\documentclass[12pt,preprint]{aastex}
\shortauthors{Belucz et al}
\shorttitle{Dynamo with multi-cell meridional circulation}

\usepackage{graphicx}
\usepackage{amsmath}

\begin{document}

\title{A BABCOCK-LEIGHTON SOLAR DYNAMO MODEL WITH MULTI-CELLULAR MERIDIONAL 
CIRCULATION IN ADVECTION- AND DIFFUSION-DOMINATED REGIMES}

\author{BERNADETT BELUCZ$^1$, MAUSUMI DIKPATI$^2$ \& EMESE FORG\'ACS-DAJKA$^1$}

\affil{1. E\"otv\"os University, Department of Astronomy, 1518 Budapest,
Pf. 32, Hungary}

\affil{2. High Altitude Observatory, National Center for Atmospheric
Research, 3080 Center Green, Boulder, CO 80307-3000.}

\email{bbelucz@astro.elte.hu; dikpati@ucar.edu}

\begin{abstract}

Babcock-Leighton type solar dynamo models with single-celled
meridional circulation are successful in reproducing many solar
cycle features. Recent observations and
theoretical models of meridional circulation do not indicate a 
single-celled flow pattern. We examine the role of complex multi-cellular
circulation patterns in a Babcock-Leighton solar dynamo in advection- and 
diffusion-dominated regimes. We show from simulations that 
presence of a weak, second, high-latitude reverse cell speeds up 
the cycle and slightly enhances the poleward branch in butterfly 
diagram, whereas the presence of a second cell in depth reverses 
the tilt of butterfly wing to an anti-solar type. A butterfly diagram 
constructed from middle of convection zone yields a solar-like 
pattern, but this may be difficult to realize in the Sun because of 
magnetic buoyancy effects. Each of the above cases behaves  
similarly in higher and lower magnetic diffusivity regimes. However, our 
dynamo with a meridional circulation containing four cells in 
latitude behaves distinctly differently in the two regimes, producing 
solar-like butterfly diagrams with fast cycles in the higher diffusivity 
regime, and complex branches in butterfly diagrams in the 
lower diffusivity regime. We also find that dynamo solutions for a 
four-celled pattern, two in radius and two in latitude, prefer to
quickly relax to quadrupolar parity if the bottom flow-speed is
strong enough, of similar order of magnitude as the surface flow-speed.

\end{abstract}



\section{INTRODUCTION}

Over the past two decades Babcock-Leighton type \citep{b61, l64, l69} solar
dynamo models operating with single celled meridional circulation have been
successful in reproducing many solar cycle features, including equatorward
migration of sunspot belts, poleward drift of poloidal fields and the correct
phase relationship between them \citep{w91, csd95, d95, dc99, krs01, berb02, 
nc01, gm04, jetal08}. It was possible to calibrate these models for the Sun and
so they were applied to prediction of solar cycle amplitude. Now we know that 
solar cycle 24 amplitude forecast of \citet{ddg06, dg06}, namely a 30\% to 50\% stronger 
cycle 24 peak than the peak of cycle 23, may not be validated. One of the
reasons is that the assumption of a steady, single-celled meridional circulation
in each hemisphere may be oversimplified for the Sun. Both observations
and models indicate that there may be more than one cell in either depth
or latitude, or both, in each hemisphere, at least at some times.

Using time-distance helioseismology the most recent observations from 
SDO/HMI data infer meridional circulation with two cells in depth 
\citep{z13}. Ring-diagram analysis from GONG data gives 
poleward surface flow up to about $\sim 60^{\circ}$ latitude \citep{h02,
ba10, ketal13}, whereas Doppler measurements from MWO data, which can measure this 
flow at higher latitudes, show a high-latitude, reverse flow associated
with the primary poleward surface flow \citep{u10}. 
Using a very long-term GONG database and applying time-distance
technique \citet{kholikov14} have found signatures of equatorward 
return-flow in the lower half of the convection zone, indicating a
long deep one cell flow-pattern. A p-mode perturbation analysis by
\citet{sch13} yields four cells in latitude, each going down to about
0.8R. Thus observations do not yet give us a unique answer about the
Sun's meridional circulation pattern.

Models produce more complex flow patterns that vary from model to model. 
For example, mean-field models produce a long, counterclockwise primary 
cell often associated with a weak, reverse cell at high latitudes, both 
extending down to the bottom of convection zone \citep{r89,kr05,dikpati14}, 
whereas full 3D convection simulations produce multiple cells in latitude 
and depth \citep{gskm13,fm15}. Recently \citet{dam14} have shown the scope
of application of Ensemble Kalman filter data assimilation in a flux-transport 
dynamo model for reconstructing the time-variation of the Sun's meridional 
circulation, and in the future the combination of theory, observation and 
data assimilation can be implemented to derive the spatio-temporal pattern 
of the Sun's meridional circulation. However, given the lack of knowledge 
about the uniqueness of meridional flow from observations and models at 
present, it is necessary to consider all plausible meridional circulation 
patterns for the Sun, and explore their effects on a Babcock-Leighton solar 
dynamo model. We specifically seek the answers to the following questions: 
(i) can a Babcock-Leighton dynamo operating with a multi-cellular meridional 
circulation pattern produce observed solar cycle features? (ii) Can such a 
model be calibrated for the Sun in advection and diffusion-dominated regimes?  

In the present paper, our aim is to study the effects of various 
plausible multi-cellular meridional flow patterns on Babcock-Leighton
solar dynamos operating in a full spherical shell of the convection 
zone. It is important to use inputs to the model, such as meridional
circulation, that are as closely guided by available observations as 
possible, to get the best possible model dynamo properties that can 
be compared with properties of observed solar cycles. Guided by the 
observational and modeling results cited above, we choose five circulation
patterns. These include in each hemisphere (a) a single cell with a poleward
flow at the surface; (b) a long primary cell from the equator to about 
$60^{\circ}$ latitude, associated with a second, reversed cell at high
latitudes; (c) two cells in depth; (d) two cells in depth and two in
latitude; and (e) four cells in latitude. 

Flux transport dynamo simulations have been done for some of these 
cases, but before the latest helioseismic observations of meridional 
circulation was available, so the choices of circulation patterns was 
less closely aligned to these observations. For example, \citet{bebr05} 
simulated a flux-transport dynamo operating with two cells in latitude
having similar latitudinal extent and amplitude. The observations clearly show
the low latitude cell always reaches to at least $60^{\circ}$ and the second
cell beyond it is quite weak compared to the primary cell (our case (b)). 
These differences lead to substantial differences in butterfly diagram,
generally in the direction of poorer agreement with solar cycle observations.

\citet{jb07} have explored a Babcock-Leighton flux-transport dynamo with 
meridional circulation patterns equivalent to our cases (a), (c) and (e).
However, there are a number of differences between our cases and theirs,
which we explain below. Their motivation in large part was to try to improve on
the success of solar flux transport dynamo models with single celled
meridional circulation in simulating solar cycle features. However, they 
generally found that two- and four-celled circulations led to butterfly diagrams
and other characteristics that were less like the Sun rather than closer to it.
In this paper we use the single celled dynamo results as the reference case, 
and focus on which other circulation patterns suggested by most recent
observations could also do as well as the reference case, and which give 
results that diverge significantly from the solar observations.

In \citet{jb07} all four cells have similar amplitudes and latitudinal
dimensions, again significantly different from recent solar observations.
In addition, the streamfunction for their meridional circulation is
physically unrealistic for the Sun, because it is computed using a density profile
that varies like $1/r$ across the convection zone. This means that the density 
at the bottom is only $50\%$ higher than at the outer boundary, so the flow is
almost incompressible. The effect is to make the deep circulation cells much 
more nearly equal in amplitude to the surface velocities than is likely to be the 
case in the Sun. We will use the same density profile as taken in \citet{dc99}, 
in which density is proportional to ${(R_{\odot}/r-\gamma)}^{m})$. This 
density profile gives a density difference between top and bottom that is 
similar to a polytrope for an adiabatically stratified solar/stellar 
convection zone, as well as to typical profiles used in helioseismology such as 
in \citet{christensen96}. For values of $\gamma \approx 1$ and $m=1.5$, the 
density near the top becomes much more like the Sun for a bottom density of 
0.2 ${\rm gm}\,{\rm cm}^{-3}$. With this density profile, the bottom cell 
becomes much smaller in amplitude than is the top cell, as found by \citet{z13} 
from SDO/HMI data. These differences in cell amplitude also strongly affect 
the butterfly diagram and surface poloidal fields.

The inference of two cells in depth by \citet{z13} have created enough
interest to explore how a flux-transport dynamo behaves with such a 
flow pattern. Being poleward at the base of the convection zone, this 
flow transports spot-producing tachocline toroidal fields poleward. Our 
case (c) will address this issue. Our case (d) is another form of two cells 
in depth, but the cells do not go all the way to the pole, instead reverse 
beyond $60^{\circ}$ latitude. We study the role of such a four-celled 
pattern (two cells in depth and two in latitude), because \citet{z13} did 
not confirm whether the poleward surface flow continues to the pole or stops 
at high latitudes. On the other hand, surface Doppler measurements indicate 
that the surface flow is poleward up to a certain high latitude 
($\sim 60^{\circ}$); beyond that, \citet{u10} found a reverse, 
equatorward flow during some epochs, although not all the time. 

We examine which of the five circulation patterns mentioned in (a-e)
can produce cyclic features similar to the Sun. Furthermore we perform 
a systematic parameter survey to compare dynamo model simulations for 
all five circulations using the same model and the same dynamo physics,
and judge which models that use these circulation patterns can be 
calibrated to the Sun in diffusion and advection dominated regimes.
After describing the model in the next section, we present our results
in \S3 and conclude in \S4.

\section{MODEL}

\subsection{Dynamo equations.}

We use the spherical polar coordinates $\mathrm{(r, \theta, \phi)}$, and 
assume the axisymmetry. The magnetic field, as the sum of a toroidal 
component ($\mathrm{B_\phi}$) and a poloidal component 
($\mathrm{\mathbf{B_p}}$), can be written as:
\begin{equation}
\mathrm{\mathbf{B}(r,\theta,t)=B_\phi(r,\theta,t)\hat{\pmb{e}}_\phi
+\boldsymbol{\nabla}\times\left[A(r,\theta,t)
\hat{\pmb{e}}_\phi\right]}
\end{equation}
where the toroidal component of the magnetic field 
$\mathrm{B_\phi(r,\theta,t)\hat{\pmb{e}}_\phi}$ and the vector potential 
$\mathrm{A(r,\theta,t)\hat{\pmb{e}}_\phi}$. The both components can be generated 
by a flow. The large-scale flow field $\mathrm{\mathbf{U}(r,\theta)}$ be expressed 
as the sum of differential rotation $(\mathrm{\Omega(r,\theta)})$ and the 
meridional circulation 
$(\mathrm{\mathbf{u}(r,\theta)}=u_r(r,\theta) \hat{\pmb{e}}_r+
u_\theta(r,\theta)\hat{\pmb{e}}_\theta)$:
\begin{equation}
\mathrm{\mathbf{U}(r,\theta)=\mathbf{u}(r,\theta)+
r\sin\theta\Omega(r,\theta)\hat{\pmb{e}}_\phi}
\end{equation}
as toroidal and poloidal parts of the total flow field. 

The evolution of the large-scale magnetic field B according to:
\begin{equation}
\mathrm{\frac{\mathbf{\partial B}}{\partial t}=
\boldsymbol{\nabla}\times(\mathbf{U}\times\mathbf{B}
-\eta\boldsymbol{\nabla}\times \mathbf{B})},
\end{equation}
The toroidal component becomes
\begin{equation}
\mathrm{\frac{\partial B_\phi}{\partial t} + \frac{1}{r} \left[\frac{\partial}
{\partial r}(r u_r B_{\phi}) + \frac{\partial}{\partial \theta}(u_{\theta}
B_{\phi}) \right] = 
r\sin\theta(\mathbf{B_p}\cdot\boldsymbol{\nabla})\Omega-
\boldsymbol{\nabla}\eta\times\boldsymbol{\nabla}\times{B_\phi \hat{\pmb{e}}_{\phi}}
}
\nonumber
\end{equation}
\begin{equation}
\mathrm{
+\eta\left(\nabla^{2}-\frac{1}{r^2\sin^{2}\theta}\right)B_\phi}
\label{eq:dBdt}
\end{equation}
where $\mathrm{\eta(r)}$ is the magnetic diffusivity.
\begin{equation}
\mathrm{\frac{\partial{A}}{\partial{t}}  
+ \frac{1}{r\sin\theta}(\mathbf{u}\cdot\boldsymbol{\nabla})
(r\sin\theta A)= \eta\left(\nabla^{2}-\frac{1}{r^2\sin^{2}\theta}\right)A
+S(r,\theta;B_\phi)}
\label{eq:dAdt}
\end{equation}
where $\mathrm{\mathbf{B_p}=\boldsymbol{\nabla}\times(A \hat{\pmb{e}}_\phi)}$. 

There is considerable uncertainty about what is the most realistic 
profile of turbulent magnetic diffusivity with radius. 
Direct measurements of magnetic diffusivity as a function of depth are not 
possible yet. The mixing-length theory gives us a rough estimate of the 
supergranular diffusivity in the supergranulation layer near the surface with 
a range ($\eta_{\mathrm{super}}=10^{12}-10^{14}\,\mathrm{cm^2s^{-1}}$). The 
magnetic diffusivity in the convective envelope of the Sun is dominated by its 
turbulent contribution, but below the convection zone there is much less 
turbulence, the core is stabler, so the diffusivity should be determined 
essentially from the molecular contribution in the stably stratified deep 
radiative interior \citep{dgm06}. We assume that the turbulence governs the 
diffusivity in the convection zone and is significantly reduced in the 
subadiabatically stratified radiative zone below \citep{ddgaw04}. 
The diffusivity profile can be written as:
\begin{equation}
\mathrm{\eta(r)=\eta_{\mathrm{core}}+
\frac{\eta_\mathrm{T}}{2}\left[1+\mathrm{erf}\left(\frac{r-r_8}{d_8}\right)\right]
+\frac{\eta_{\mathrm{super}}}{2}\left[1+\mathrm{erf}\left(\frac{r-r_9}{d_9}\right)
\right]}
\label{eqn:eta}
\end{equation}
The parameters in the diffusivity profile: $\eta_\mathrm{T}=
3\cdot{10}^{11}\,\mathrm{cm^2s^{-1}}$ is the turbulent
diffusivity, $\eta_\mathrm{core}=10^9\,\mathrm{cm^2s^{-1}}$ is the core diffusivity, 
$\eta_\mathrm{super}=3\cdot 10^{12}\,\mathrm{cm^2s^{-1}}$ is the supergranular 
diffusivity, $\mathrm{r_8}=0.7\,\mathrm{R_\odot}$, $\mathrm{d_8}=
0.0125\,\mathrm{R_\odot}$, $\mathrm{r_9}=0.9562\,\mathrm{R_\odot}$, 
$\mathrm{d_9}=0.025\,\mathrm{R_\odot}$.

A two-step profile, which leaves out the supergranule layer, will have impact on 
the surface poloidal fields that are used to create the poloidal part of the 
butterfly diagram.

Our flux-transport dynamo can be driven by both a tachocline $\alpha$-effect and 
Babcock-Leighton type surface poloidal source. The decay of tilted
bipolar active regions produce poloidal fields near the surface, but perhaps that 
cannot be the sole driver for the large-scale solar dynamo. A Babcock-Leighton 
dynamo is not a self-excited dynamo, and hence cannot come back to a normally
cycling dynamo if it goes to a grand minimum state. Flux-transport solar dynamo 
models with both Babcock-Leighton type surface poloidal source and convection 
zone $\alpha$-effect have been used \citep{pnh14, hpn14}, in the 
context of reviving a Babcock-Leighton dynamo from a grand minima. 
Furthermore, it is well known that a flux transport dynamo driven solely by a 
Babcock-Leighton poloidal source cannot sustain its antisymmetric magnetic field 
about the equator, as inferred from Hale's polarity rule \citep{henj19}, but 
slowly drifts to symmetric magnetic field solutions. This is true even in the case
of a 3D Babcock-Leighton dynamo (see, e.g. \citet{md14}). Inclusion of a tachocline 
instability-driven $\alpha$-effect is demonstrated to be one of the solutions 
to stop such a drift to a nonsolar-like, quadrupolar parity in a pure 
Babcock-Leighton dynamo models \citep{dg01, berb02}. In the present paper,
since one of our goals is to investigate the parity produced by a Babcock-Leighton
dynamo model operating with different multi-cellular meridional circulation
patterns, we assumed zero tachocline $\alpha$-effect in all cases. 
We use the following expressions respectively for Babcock-Leighton surface 
source:

\begin{equation}
\mathrm{S(r,\theta,B_\phi)=\left\{S_{\mathrm{BL}}(r,\theta)+
S_{\mathrm{tachocline}}(r,\theta)\right\}
\left[1+\left(\frac{B_{\phi}(r,\theta,t)}{B_0}\right)^2\right]^{-1}}
\end{equation} 

\begin{equation}
S_{\mathrm{tachocline}}(r,\theta)=0 
\end{equation}

\begin{equation}
\mathrm{S_{\mathrm{BL}}(r,\theta)=\frac{s_1}{2}\left[1+\mathrm{erf}
\left(\frac{r-r_4}{d_4}\right)\right]
\left[1-\mathrm{erf}\left(\frac{r-r_5}{d_5}\right)\right]\sin\theta\cos\theta
\left[\frac{1}{1+e^{\gamma_1(\pi/4-\theta)}}\right]}
\label{eqn:source1}
\end{equation} 

for $\theta<\pi/2$ and for $\theta>\pi/2$
\begin{equation}
\mathrm{S_{\mathrm{BL}}(r,\theta)=\frac{s_1}{2}\left[1+\mathrm{erf}
\left(\frac{r-r_4}{d_4}\right)\right]
\left[1-\mathrm{erf}\left(\frac{r-r_5} {d_5}\right)\right]\sin\theta\cos\theta
\left[\frac{1}{1+e^{\gamma_1(\theta-3\pi/4)}}\right]}
\label{eqn:source2}
\end{equation} 

The parameters used in \eqref{eqn:source1} and \eqref{eqn:source2} are: 
$\mathrm{s_1}=2.5\,\mathrm{ms^{-1}}$, $\mathrm{r_4}=0.95\,\mathrm{R_\odot}$, 
$\mathrm{r_5}=0.9875\,\mathrm{R_\odot}$, $\mathrm{d_4=d_5}=0.0125\, \mathrm{R_\odot}$, 
$\mathrm{\gamma_1}=30.0$, $\beta=70.0$, and $\mathrm{B_0}=10\,\mathrm{kG}$.

The solar internal rotation profile includes primarily latitudinal shear in the 
convection zone, as found by two-dimensional helioseismic inversions. A solar-like 
internal differential rotation profile (\cite{dc99}) is given by \begin{equation}
\mathrm{\Omega(r,\theta)=\Omega_c+\frac{1}{2}\left[1+\mathrm{erf}
\left(2\frac{r-r_c}{d_1}\right)\right]\{\Omega_s (\theta)-\Omega_c\}}
\end{equation}
where
\begin{equation}
\mathrm{\Omega_s=\Omega_{Eq}+a_2{\cos^{2}\theta}+a_4{\cos^{4}\theta}}
\end{equation}
is the surface latitudinal differential rotation. Values were chosen to closely 
resemble the best fit to the helioseismic solution of \cite{ctst98}. The angular 
velocity of the rigidly rotating core is $\mathrm{\Omega_c/2\pi=432.8\,nHz}$. 
$\mathrm{\Omega_{Eq}/2\pi=460.7\,nHz}$ is the rotation rate at the equator. The 
other parameters are set to be $\mathrm{a_2/2\pi=62.69\,nHz}$, and 
$\mathrm{a_4/2\pi=67.13\,nHz}$. The $\mathrm{r_c=0.7\,R_{\odot}}$ indicates the 
central radius of the tachocline thickness $\mathrm{d_1=0.025\,R_{\odot}}$. 

The differential rotation profile prescribed by three terms, as shown in
Equation (12), is well-formulated up to $60^{\circ}$, but to adequately
fit the measured rotation rate including more poleward latitudes requires
additional terms \citep{schou98, dctg02}. We will discuss how the implementation
of such differential rotation profile affects our results. 

\subsection{Stream function}

The circulation is represented in the spherical shell by the stream function of 
\citet{fd02} for both hemispheres. To study the effects of the meridional 
circulation, we use a simple and easily adjustable spatial structure, with 
realistic amplitude. The components of the meridional circulation can be written as
\begin{equation}
{u_r}(r,\theta)=\frac{1}{\rho(r)\cdot r^2 \sin\theta}\cdot
\frac{\partial}{\partial\theta}\mathbf{\Psi}(r,\theta)
\end{equation}
\begin{equation}
{u_{\theta}}(r,\theta)=\frac{-1}{\rho(r)r\sin\theta}\cdot
\frac{\partial}{\partial{r}}\mathbf{\Psi}(r,\theta)
\end{equation}
Note that the spherical polar geometrical factor ($r\,\sin\theta$) is
absorbed in $\mathbf{\Psi}(r,\theta)$.

\begin{figure}[hbt]
\epsscale{1.0}
\plotone{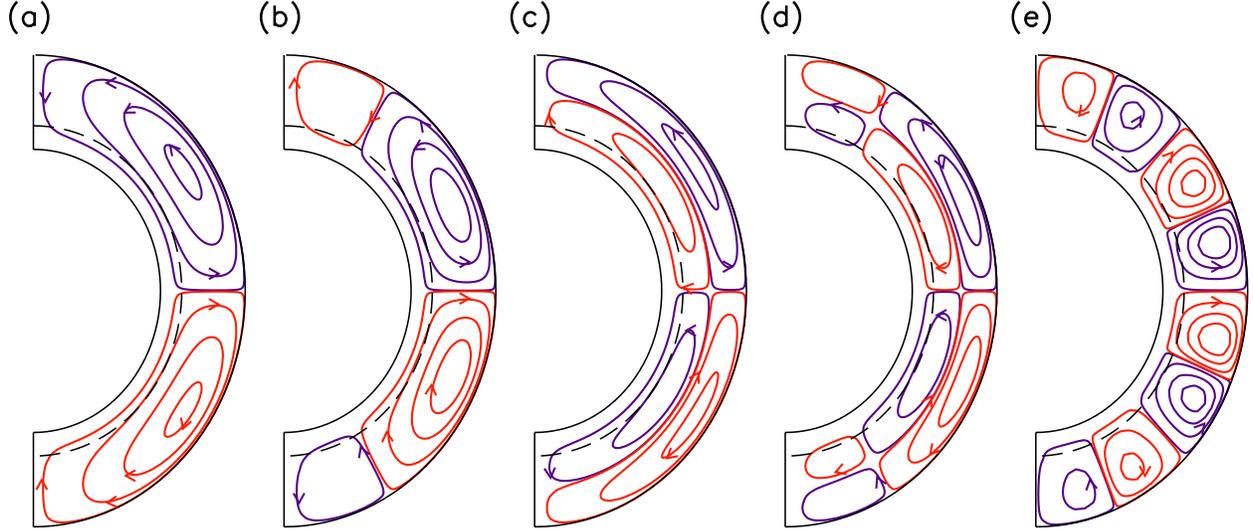}
\caption{Streamlines of the meridional circulation as prescribed in 
\eqref{eqn:sf2}.}
\label{streamf}
\end{figure}

We reproduce the stream function of \citet{fd02}, which has the following form:
\begin{equation}
\mathrm{\mathbf{\Psi}(r,\theta)=\mathbf{\psi}(r)\sin^2 n_1\theta\cos n_2\theta}
\end{equation}
where the form of the given function $\psi(r)$ specifies the flow.
\begin{equation}
\mathrm{\psi(r)=\psi_0\sin\left[\frac{k\pi\left(r-r_{mc}\right)}{R_{\odot}-r_{mc}}\right]
\exp\left[\frac{(r-r_0)^2}{\Gamma^2}\right]}
\label{eqn:sf2}
\end{equation}
where $\mathrm{\psi_0}$ sets the amplitude of the meridional circulation. Two 
parameters play the primary role in setting the meridional circulation for the 
simulations. We fix the number of cells in latitude by changing the parameter $n$
and the number of cells in radius by changing the parameter $k$. The $\mathrm{r_0}$ 
and $\mathrm{\Gamma}$ are  geometric parameters, $\mathrm{r_0=(R_{\odot}-r_{mc})/30\,cm}$ 
and $\mathrm{\Gamma=6.4\cdot 10^{10}\,cm}$. $\mathrm{r_{mc}}$ is the radius to 
which the meridional flow penetrates from the base of the convective zone. Given 
the observed depth of the tachocline, we set $\mathrm{r_{mc}=4.79\cdot 10^{10}\,cm}$. 
We use the same density profile ($\rho(r)$) as used by \citet{dc99}.

With all parameters specified, we can compute the stream functions for the five 
circulation patterns we use in the dynamo. Figure 1 displays the circulation 
patterns we will use for our analysis of the differences in dynamo behavior that 
arise from different circulation patterns. These include the reference case (a) 
a single cell in each hemisphere; (b) a long, primary cell associated with a weak 
reverse cell in high latitudes; (c) two cells with depth; (d) two cells in 
latitude and two in depth; and (e), four cells in latitude. There is some 
observational support for each of these patterns, as reviewed in the introduction.
The parameters chosen in the five cases are given respectively as: (a) $k=1$,
$n_1=n_2=1$, $\Psi_0=-6.1771 \times 10^{21}$; (b) $k=1$, $n_1=1$, $n_2=3$,
$\Psi_0=3.519 \times 10^{21}$; (c) $k=2$, $n_1=1$, $n_2=1$, $\Psi_0=3.0937 
\times 10^{21}$; (d) $k=2$, $n_1=1$, $n_2=3$, $\Psi_0=-1.7625 \times 10^{21}$;
(e) $k=1$, $n_1=4$, $n_2=4$, and $\Psi_0= {Psi_0}_a + {\Psi_0}_b + {\Psi_0}_c
+ {\Psi_0}_d$, in which we assign values for ${\Psi_0}_a, {\Psi_0}_b, {\Psi_0}_c, 
{\Psi_0}_d$ for a certain range in $\theta$ and zero elsewhere, namely
${\Psi_0}_a = 1.7043 \times 10^{21}$ for $0\le \theta \le22.5^{\circ}$,
${\Psi_0}_b = 4.1340 \times 10^{21}$ for $22.5^{\circ} \le \theta \le45^{\circ}$,
${\Psi_0}_c = -6.8972 \times 10^{21}$ for $45^{\circ} \le \theta \le 67.5^{\circ}$,
${\Psi_0}_d = -7.8002 \times 10^{21}$ for $67.5^{\circ} \le \theta \le 90^{\circ}$.
Unit of $\Psi_0$ is in c.g.s, i.e., ${\rm cm}^3\,{\rm s}^{-1}\,{{\rm unit \, of} \,
 \rho}^{-1}$.

\subsection{Boundary conditions and solution method}

Equations \eqref{eq:dBdt} and \eqref{eq:dAdt} are solved in a full-spherical shell 
($\mathrm{0\ge \theta\ge \pi}$), extending radially from below the bottom of 
convection zone ($\mathrm{r/R_\odot=0.6}$) to the surface ($\mathrm{r/R_\odot=1}$). 
We use a 4th order Runge-Kutta method with the central finite difference scheme 
for $r$ and $\theta$ derivatives to solve the equations. Validity test of the
scheme was performed by reproducing the results the benchmark cases of a
Babcock-Leighton flux-transport dynamo \citep{jetal08}. We use the boundary 
conditions as used previously by \citet{dc99}. We set $\mathrm{B_\phi}=0$ and 
$\mathrm{A}=0$ at the bottom boundary ($r/R_{\odot}=0.6$). Note that the bottom 
boundary condition for the toroidal field is modified from the condition for 
a perfectly conducting core (see e.g. \citet{dg01} for physical explanation). 
Both $\mathrm{B_\phi}=0$ and $\mathrm{A}=0$ are also set to zero along the polar 
axis ($\theta=0\text{ or }\pi$) to ensure physicality. At the surface, the toroidal 
fields are set to zero, and we demand that the poloidal field-lines match 
smoothly with the potential field solutions of the free space above the surface
(see \citet{dc94}). The vector potential $\mathrm{A}$ satisfies the following 
equation above the surface:
\begin{equation}
\left(\nabla^2-\frac{1}{r^2 \sin^2\theta}\right)A=0.
\end{equation}
A general solution for this equation can be written as:
\begin{equation}
A(r\geq R_{\odot},\theta,t)=\sum_n\frac{a_n(t)}{r^{n+1}}P_n^l(\cos\theta)
\end{equation}
\begin{equation}
a_n(t)=\frac{(2n+1)R_{\odot}^{n+1}}{n(n+1)}\int_0^{\pi/2}A(r=R_{\odot},\theta,t)
P_n^l(\cos\theta)\sin\theta d\theta
\end{equation}
where $P_n^l(\cos\theta)$ is the associated Legendre polynomial. The derivative of 
$\mathrm{A}$ at the solar surface 
\begin{equation}
\left.\frac{\partial{A}}{\partial{r}}\right|_{(r=R_{\odot})}=
-\sum_n\frac{(n+1)a_n(t)}{R_{\odot}^{n+2}}P_n^l(\cos\theta)
\end{equation}

\section{RESULTS}

\subsection{Time-latitude diagrams for multi-cell flow}

\begin{figure}[hbt!]
\epsscale{1.0}
\plotone{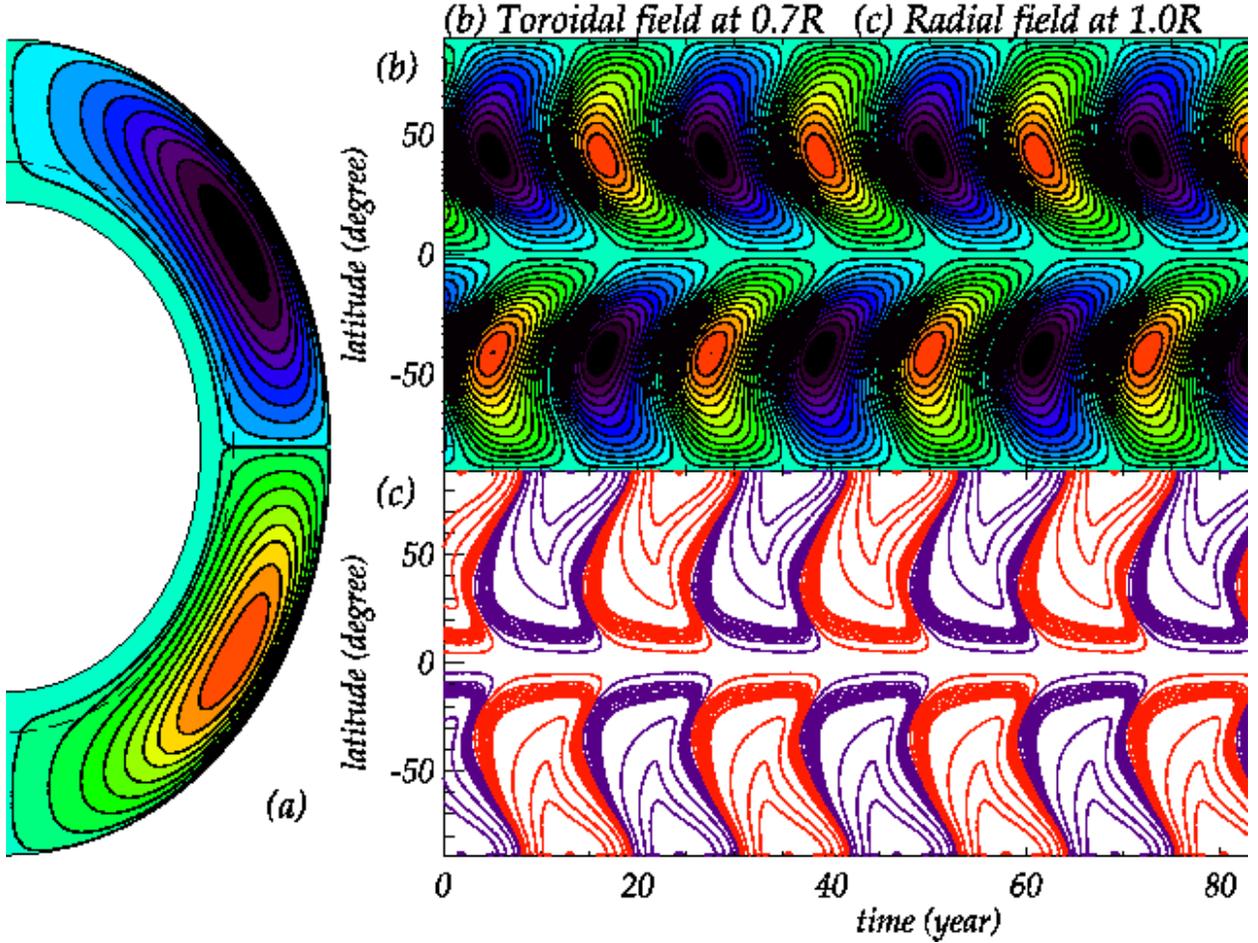}
\caption{Panel (a) displays the streamlines for single-celled meridional circulation 
in each hemisphere; blue-violet represents counterclockwise flow, green-orange 
clockwise flow.  Panel(b) shows the time evolution of the tachocline toroidal field; 
panel (c) the same for surface radial fields. For tachocline toroidal fields, 
color-filled contour levels are 3kG; the highest tachocline toroidal field is 
$\mathrm{\sim 37\,kG}$ (yellow/violet). The maximum value of the radial fields is 
$\mathrm{\pm 204\,G}$, occurring near the poles.} 
\label{AAref}
\end{figure}

First we establish a reference case dynamo solution, to which other individual solutions 
can be compared, to see what changes are created by changing the meridional circulation 
pattern. We choose for reference the frequently used single cell meridional circulation 
that has poleward flow near the outer boundary, and return flow at the base of the 
convection zone (see Figure 1(a) and Figure 2(a)). This flow penetrates slightly below 
the base of the convection zone, given by $r=0.7R_{\odot}$. The thin black dashed 
semicircular arc represents the location of the center of the tachocline. Two frames, 
(b) and (c) on the right panel of Figure \ref{AAref}, show the time-latitude diagrams 
respectively of toroidal field at the bottom of convection zone 
($\mathrm{B_\phi |_{r=0.7R_{\odot}}}$) and surface radial field 
($\mathrm{B_r |_{r=R_{\odot}}}$). To get these results we took a maximum surface 
flow speed of $\mathrm{15\,ms^{-1}}$, poloidal source amplitude of 
$\mathrm{s_1}=3.0\,\mathrm{ms^{-1}}$ and turbulent diffusivity 
$\eta_\mathrm{T}=3\cdot{10}^{11}\,\mathrm{cm^2s^{-1}}$.

\begin{figure}[hbt!]
\epsscale{1.0}
\plotone{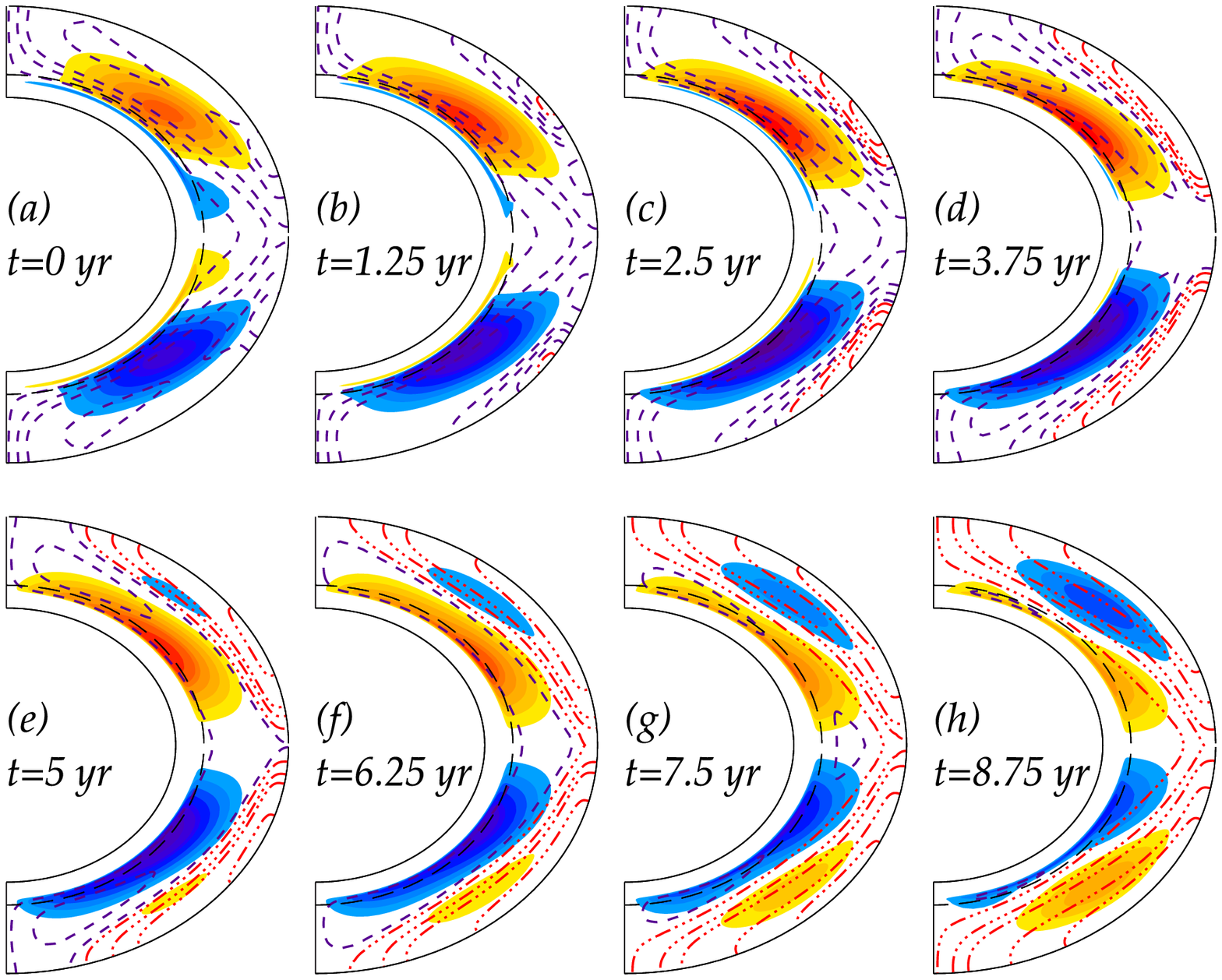}
\caption{Evolution of toroidal and poloidal fields with one-cell meridional 
circulation (Figure \ref{streamf}(a)). The filled contours show toroidal fields, 
yellow/red colors denoting positive (into the plane of the paper) and 
blue/purple negative (out of the plane). Red and blue contours respectively
denote positive (clockwise) and negative (anticlockwise) poloidal
fieldlines.}
\label{AAevolution}
\end{figure}

Figure \ref{AAevolution} shows how the poloidal and toroidal fields evolve through a
sunspot cycle for the entire meridional cross-section. When the toroidal field is 
strong, it is confined to the lower layers of the domain, and its peak is clearly 
migrating toward the equator, along with the poloidal field lines that are sheared 
by differential rotation to produce it. We expect sunspots to emerge from at or near 
the latitude of maximum in the toroidal field. The reversal of polar fields is seen 
here in the interval between 5 and 6.25 years, at which time the toroidal is a 
maximum with its peak near $20^{\circ}$. 

Babcock-Leighton flux transport dynamo models with one-celled meridional
circulation in each hemisphere can reproduce many features of the solar cycle. 
These include (i) the equatorward migration of toroidal flux at lower latitudes; 
(ii) the 11-year sunspot cycle; (iii) the observed phase-shift between poloidal 
and toroidal components; (iv) the short rise of toroidal field to maximum followed 
by the long decline to minimum -- in the reference case, the ascending phase is 
$\mathrm{16.77\%}$ of the whole cycle; (v) peak tachocline toroidal fields are 
$\mathrm{37\,kG}$; (vi) peak surface radial fields are $\mathrm{\pm 204\,G}$, 
similar to values obtained by many previous authors \citep{d95, dc99, dg01, 
rdm05, dg06, jb07, derosa05, dgtu10, bd13}. This solution also reproduces the 
observed phase shift between the surface poloidal field and the toroidal field at
the tachocline; the poloidal field polarity changes from positive to negative 
when the toroidal field is near maximum and its polarity is negative.  

\begin{figure}[hbt!]
\epsscale{1.0}
\plotone{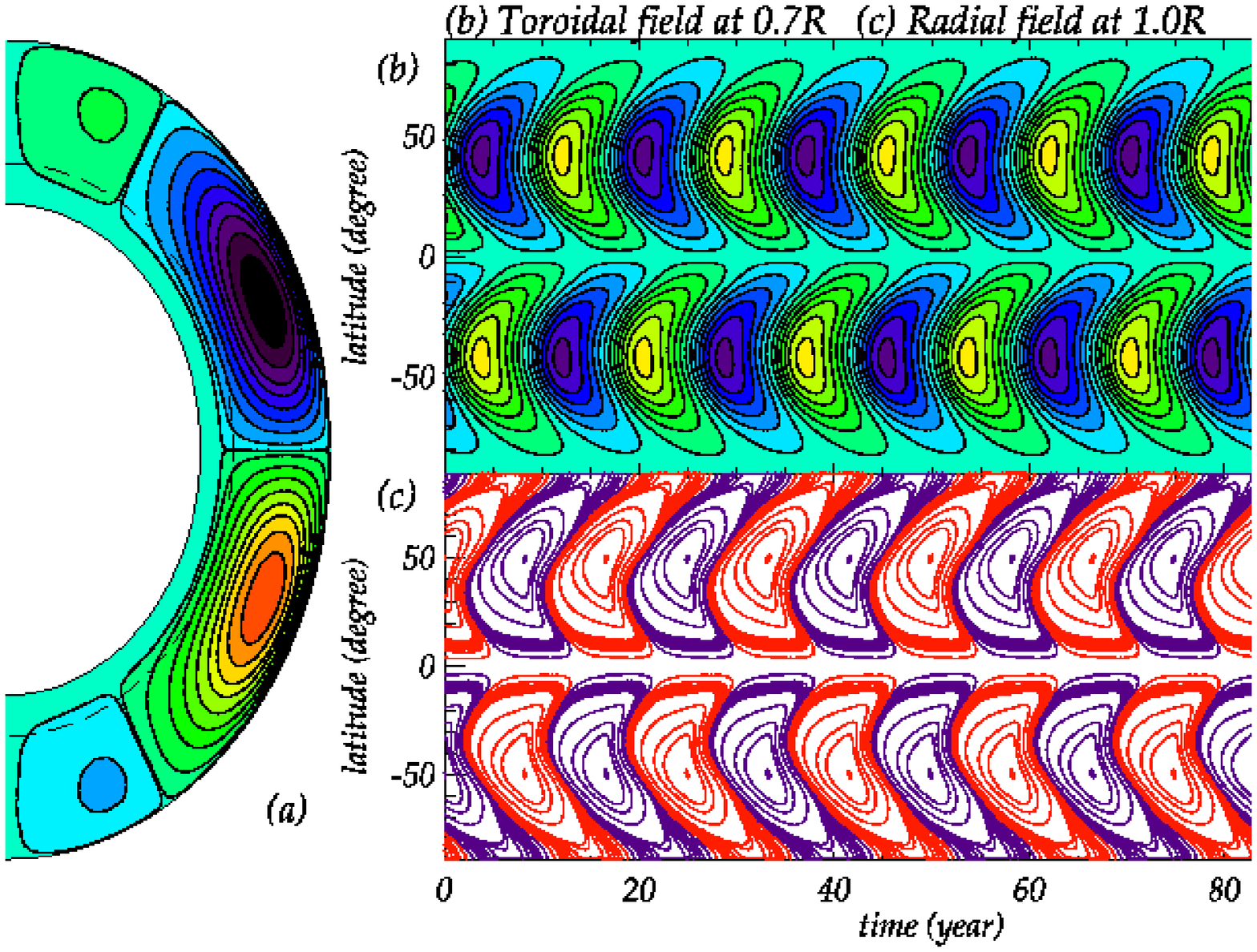}
\caption{Same as in Figure \ref{AAref} but the meridional flow pattern has 
a high-latitude second reverse flow-cell (Figure \ref{streamf}b).  
The highest tachocline toroidal field is similar to that in Figure \ref{AAref}, 
$\mathrm{\sim 37\,kG}$ (yellow/violet). The maximum value of the radial 
fields is $\mathrm{\pm 89\,G}$, at about $50^{\circ}$}.
\label{HHref}
\end{figure}

In the next simulations we study how the characteristic features of 
butterfly-diagram change when the meridional circulation cell contains a 
second, high-latitude, reverse cell. The form of this meridional circulation 
is shown in Figure 1(b) and Figure \ref{HHref}a. The peak flow-speed of the 
primary cell is still $\mathrm{15\,ms^{-1}}$, poleward at the surface, but 
the latitude of this peak is slightly lower, at $\mathrm{25.3^\circ}$. 
The peak flow speed of the secondary cell is $\mathrm{3\,ms^{-1}}$, 
equatorward at the surface. The boundary between cells is near 
$\mathrm{61^\circ}$ latitude.

The right panel of Figure \ref{HHref} shows the time-latitude diagrams of 
$\mathrm{B_\phi |_{r=0.7R_{\odot}}}$ in panel (b) and 
$\mathrm{B_r |_{r=R_{\odot}}}$ in panel (c). Not surprisingly, due to the
effect of the second cell, a more pronounced poleward branch can be seen 
in the butterfly diagram of toroidal field in panel (b) compared to that 
in Figure \ref{AAref}. The sunspot cycle length (i.e. half magnetic cycle) 
is just $\mathrm{8.3}$ years, due to the shorter primary conveyor belt. 
The strength of toroidal field is similar to that of reference case, 
$\mathrm{37\,kG}$. Comparing the time-latitude diagrams of Figure 
\ref{AAref}(c) and \ref{HHref}(c), we find that the polar field peaks around
$\mathrm{50^\circ}$ latitude instead of peaking near the pole as in the case
of a single cell in each hemisphere. This is due to flow convergence at
$\mathrm{61^\circ}$ latitude instead of at the pole. The second cell also
causes a delay in the polarity change by advecting polar fields away from the pole. 

The rise of the cycle from minimum to maximum in this case is slightly
longer compared to that in the single cell case. This is probably because 
some of the poloidal flux advected to the bottom in between the primary and 
secondary cells is advected toward the poles, retarding the early production 
of the equatorward migrating branch of toroidal field there.

\begin{figure}[hbt!]
\epsscale{1.0}
\plotone{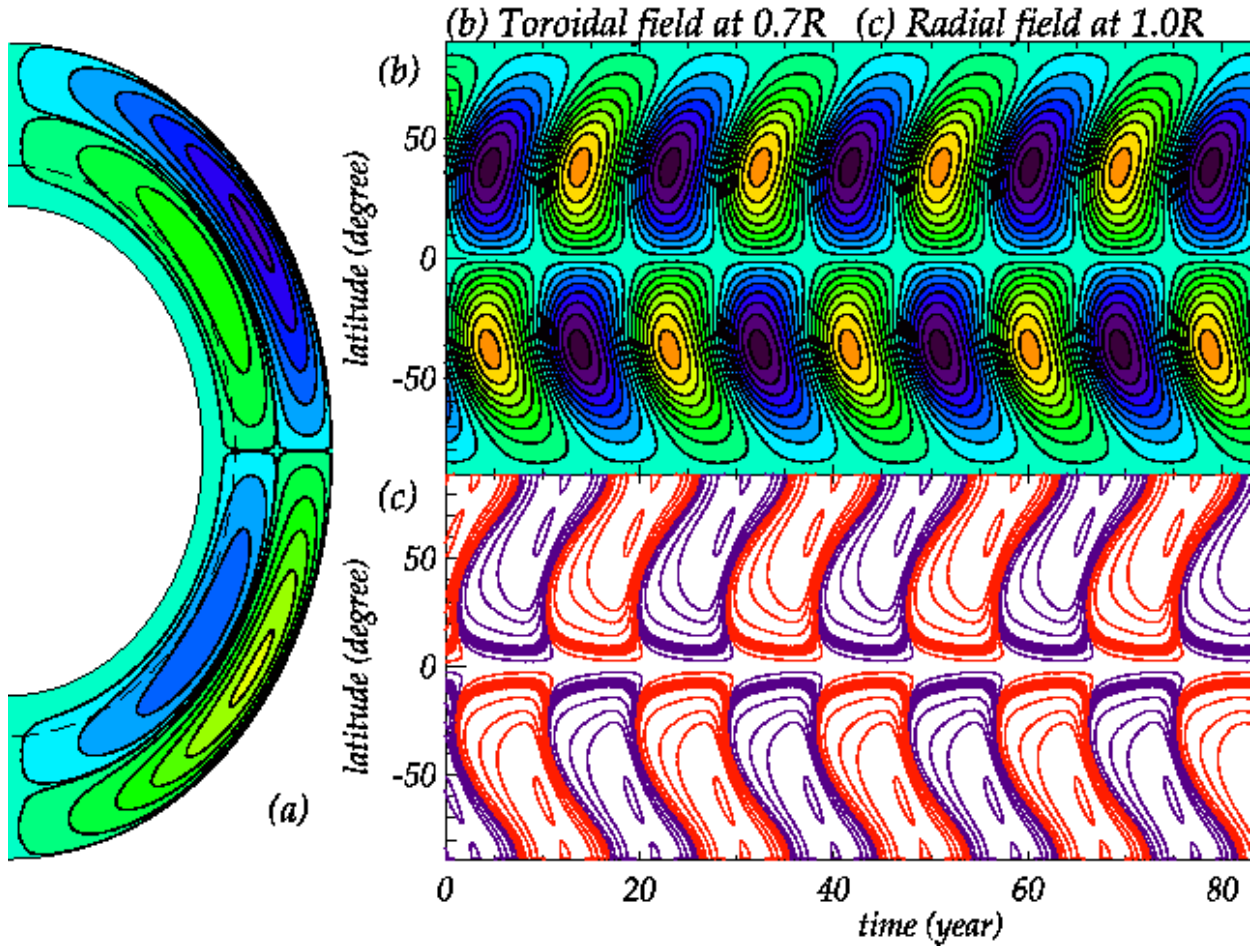}
\caption{
Same as in Figure \ref{AAref} but the meridional flow pattern has a second, 
reversed flow-cell below the primary cell (Figure \ref{streamf}c).  
The maximum tachocline toroidal field strength is $\sim 23$kG (orange/violet). 
Maximum value of the radial fields is $\pm 83$G, near the poles.}
\label{SDOSDOref}
\end{figure}

For the next simulation, we add a second, reversed meridional cell below the 
primary cell (see Figure \ref{streamf}c). The two cells are similar in 
amplitude and radial extent. The results for this meridional flow are shown 
in Figure \ref{SDOSDOref}.

\clearpage

This dynamo solution is very different from the reference case. At all 
latitudes, the tachocline toroidal field is migrating with time toward 
the poles rather than the equator. The rate of this migration increases 
with latitude. This is due to the poleward flow at the bottom of the 
second cell. Thus, based on tachocline toroidal fields, this meridional 
circulation pattern produces an 'antisolar' butterfly diagram. The toroidal 
field at the bottom of the domain is weaker than in the reference case, 
because polar fields are not brought down from the top, but instead are 
advected from lower latitudes near the bottom. 

\begin{figure}[hbt!]
\epsscale{1.0}
\plotone{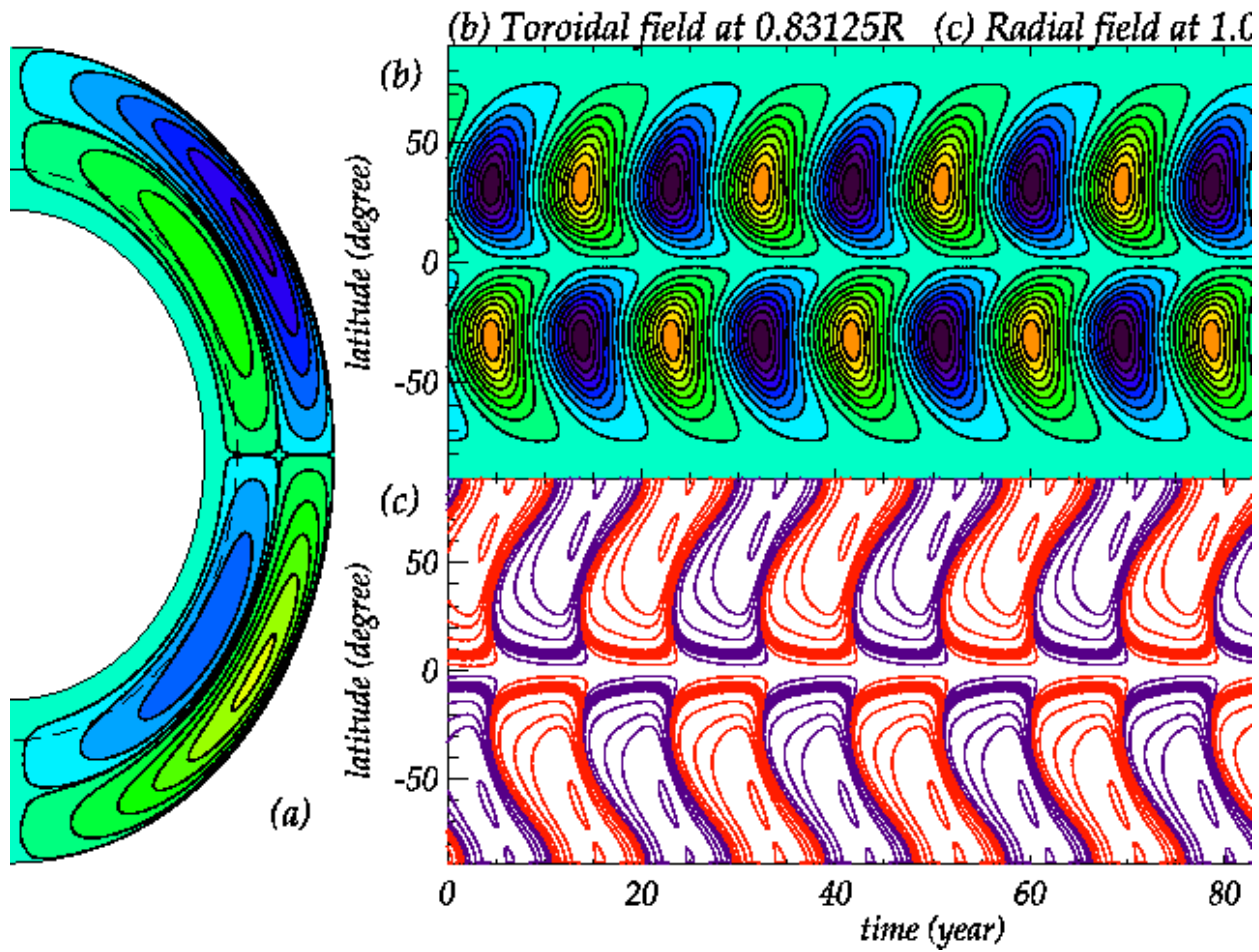}
\caption{Same as in Figure \ref{SDOSDOref} but the toroidal fields are 
shown from $0.83125\,{\rm R}$. A solar-like butterfly diagram is obtained 
to equatorward flow there.}
\label{SDOSDOcenterref}
\end{figure}

If we plot instead the toroidal field near the middle depth, for example, 
$\mathrm{0.83125\,R}$, shown in Figure 6, we get a more solar-like butterfly, 
with both poleward and equatorward branches. This is because at these 
depths both circulation cells have equatorward flow, so they advect 
toroidal field toward the equator in lower latitudes. The relatively 
high speed total flow there also makes the dynamo period slightly shorter. 
In addition, since in this case the poleward flow near the outer boundary 
reaches to a shallower depth, less poloidal flux is advected toward to 
the pole. These weaker polar fields lead to weaker toroidal fields at all
latitudes.  

\begin{figure}[hbt!]
\epsscale{1.0}
\plotone{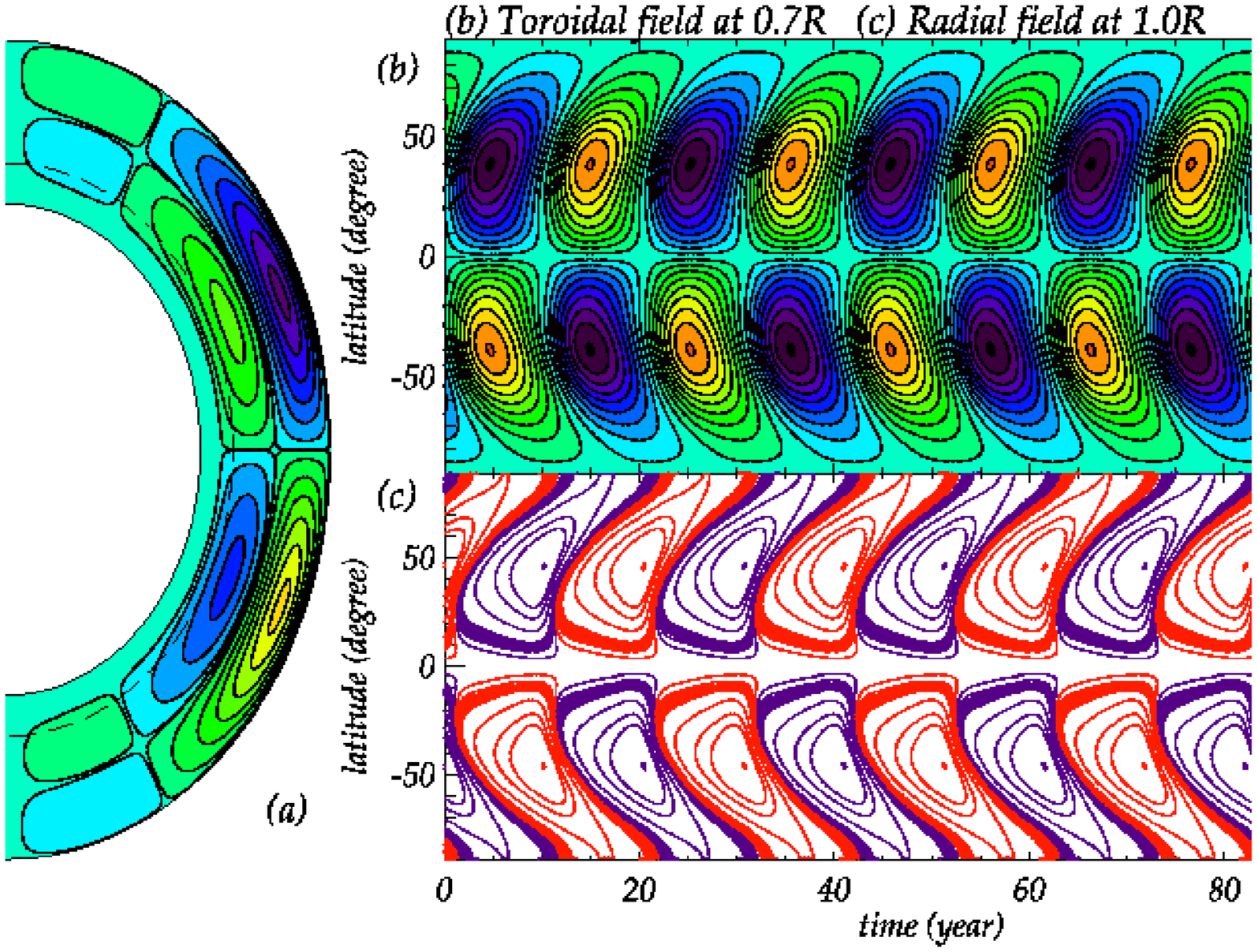}
\caption{ 
Same as in Figure \ref{AAref} but the meridional flow pattern has four 
flow-cells (Figure \ref{streamf}d).  The highest tachocline toroidal 
field is $\mathrm{\sim 33\,kG}$ (yellow/violet). The maximum value of 
the radial fields is $\mathrm{\pm 100\,G}$, at around $50^{\circ}$.}.
\label{SDOHHSDOref}
\end{figure}

The next dynamo simulation is for the case of two meridional cells in 
both latitude and depth in each hemisphere. The amplitudes of upper and lower
cells is about the same. These results are shown in Figure \ref{SDOHHSDOref}. 
Panel (b) again shows the toroidal fields near the bottom of the convective zone. 
Panel (c) depicts the surface radial field. As in the case of two cells in 
radius, we again get an 'anti-solar' butterfly diagram when we plot the toroidal
field at the bottom. Again this is because the flow is
toward the pole at the bottom of the stack of cells at all latitudes except the highest.
If we plot the toroidal field contours at mid-depth (not shown), we will again get a solar 
type butterfly, as seen in Figure 6. 

\begin{figure}[hbt!]
\epsscale{1.0}
\plotone{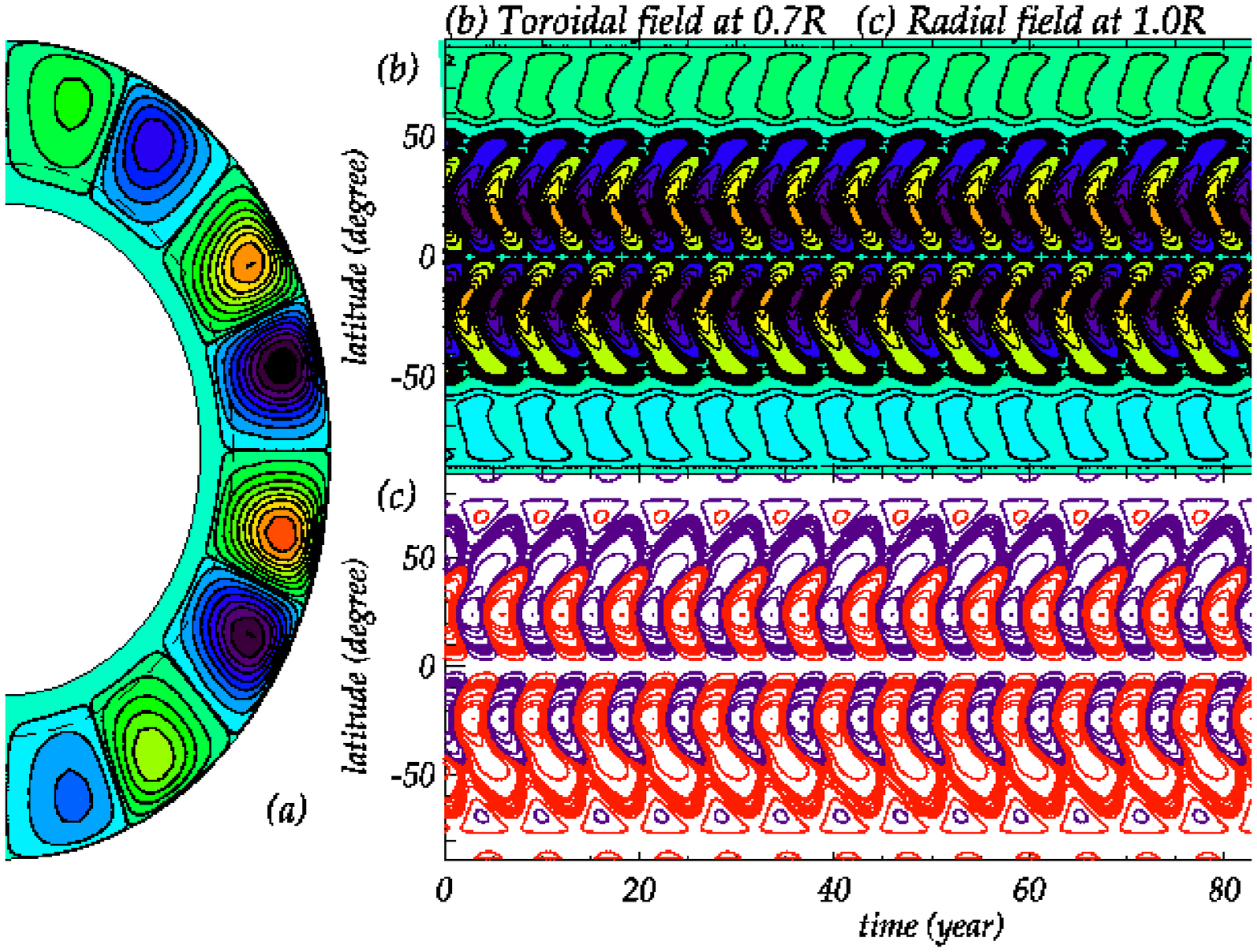}
\caption{Same as in Figure \ref{AAref} but the meridional flow pattern has 
four flow-cells (Figure \ref{streamf}e). The highest tachocline 
toroidal field is $\mathrm{\sim 15\,kG}$ (red/violet). The maximum value 
of the radial fields is $\mathrm{\pm 33\,G}$, which occurs near $25^{\circ}$.}
\label{OOref}
\end{figure}

In the last simulation, we  also have four cells, but these cells are 
located side by side, as seen in the Figure \ref{streamf}e. In this 
case, the fields are confined to lower latitudes because the multiple 
cells in latitude prevent poloidal field transport all the way to the 
poles as in the reference case. We get a solar-like time-latitude diagram 
up to about $\mathrm{22^\circ}$; the toroidal field migration is 
equatorward. We do not see dynamo activity beyond about $\mathrm{50^\circ}$. 
The cycle length is very short, 3.125 years, due to the very short
conveyor belts represented by the two circulation cells closest to the 
equator. The strength of the toroidal field is just half of the reference 
case, because the dynamo is confined to the lower latitudes where the 
differential rotation is smaller so the production of toroidal field from 
a given poloidal field is smaller.

\clearpage

\subsection{Time-latitude diagrams for multi-cell flow with lower magnetic 
diffusivity}

\begin{figure}[hbt!]
\epsscale{1.0}
\plotone{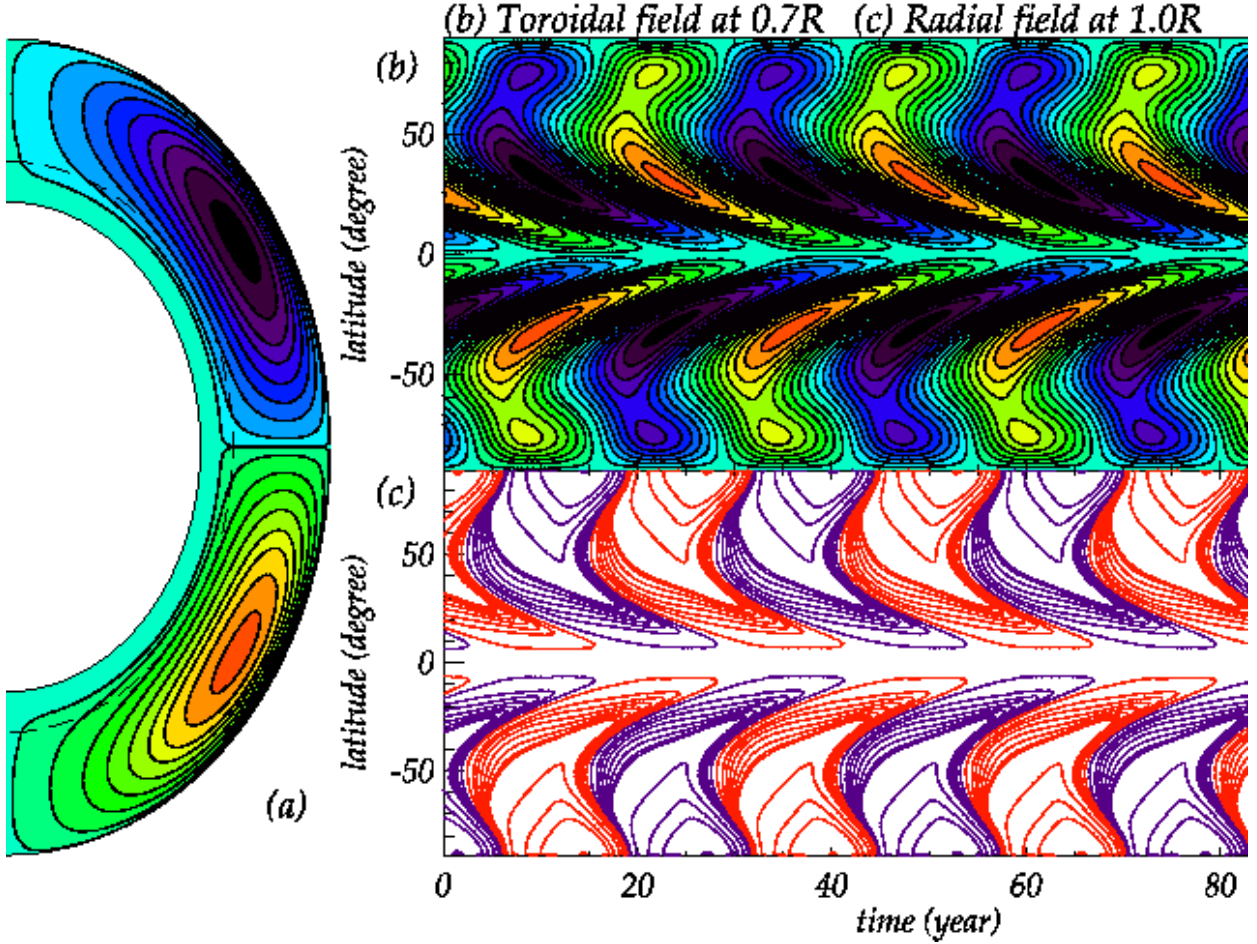}
\caption{Same as in Figure \ref{AAref} but the turbulent diffusivity is 
$\mathrm{7\cdot 10^{10}\,cm^2s^{-1}}$. The highest tachocline toroidal field is 
$\mathrm{52.4\,kG}$ (red/violet). The maximum value of the radial fields 
is $\mathrm{\pm 207\,G}$ near the poles.}
\label{AAloweta}
\end{figure}

How dependent are the results described above on the particular choice of 
turbulent magnetic diffusivity? We address that question by repeating the 
dynamo simulations for the five meridional circulations using a lower value,
$7\cdot 10^{10}\,{\rm cm}^2 {\rm s}^{-1}$. We selected this value of 
diffusivity ($\eta_T= {7\cdot 10^{10}\, {\rm cm}^2 {\rm s}^{-1}}$) to 
represent the low-diffusivity cases, and in the previous subsection
\S3.1 a diffusivity value of ${3\cdot 10^{11}\, {\rm cm}^2 {\rm s}^{-1}}$
to denote the representative value for high diffusivities. However, 
these values are not unique; neighboring values can be considered also.

The results for low diffusivity are displayed in 
Figures 9-13. Figure \ref{AAloweta} displays the reference case with 
single circulation cell for the lower magnetic diffusivity. As in earlier 
figures, panel (b) gives the toroidal field amplitude near the bottom, 
panel (c) the surface poloidal field. We use the same peak surface flow 
speed ($\mathrm{15\,ms^{-1}}$) and poloidal source amplitude 
($\mathrm{s_1}=3.0\,\mathrm{ms^{-1}}$) as for the solution seen in Figure 2.

In this lower diffusivity case, the model still reproduces the observed 
phase shift between the surface poloidal field and the tachocline toroidal 
field. Due to the lower turbulent diffusivity, much stronger fields 
are produced. The highest tachocline toroidal field is $\mathrm{52\,kG}$, 
and the maximum value of the radial fields is $\mathrm{\pm 207\,G}$ near 
the poles. The period is also somewhat longer than in Figure 2, namely 
$\mathrm{12.7}$ years, as measured by the time between adjacent peaks 
in toroidal and poloidal field at the same latitude. But there is also 
significant overlap between adjacent cycles, so the time between the 
high latitude peak of a new sign of toroidal field to its disappearance 
at low latitudes is more than $\mathrm{20}$ years. How much overlap there 
is in the Sun itself is unclear, since from observations we know where 
the tachocline toroidal field is only from the latitude of sunspots seen. 
The slight poleward migration of tachocline toroidal fields seen at high 
latitudes comes from the strong negative radial shear there overcoming 
the relatively weak equatorward advection of toroidal field by the meridional
flow there.

\begin{figure}[hbt!]
\epsscale{1.0}
\plotone{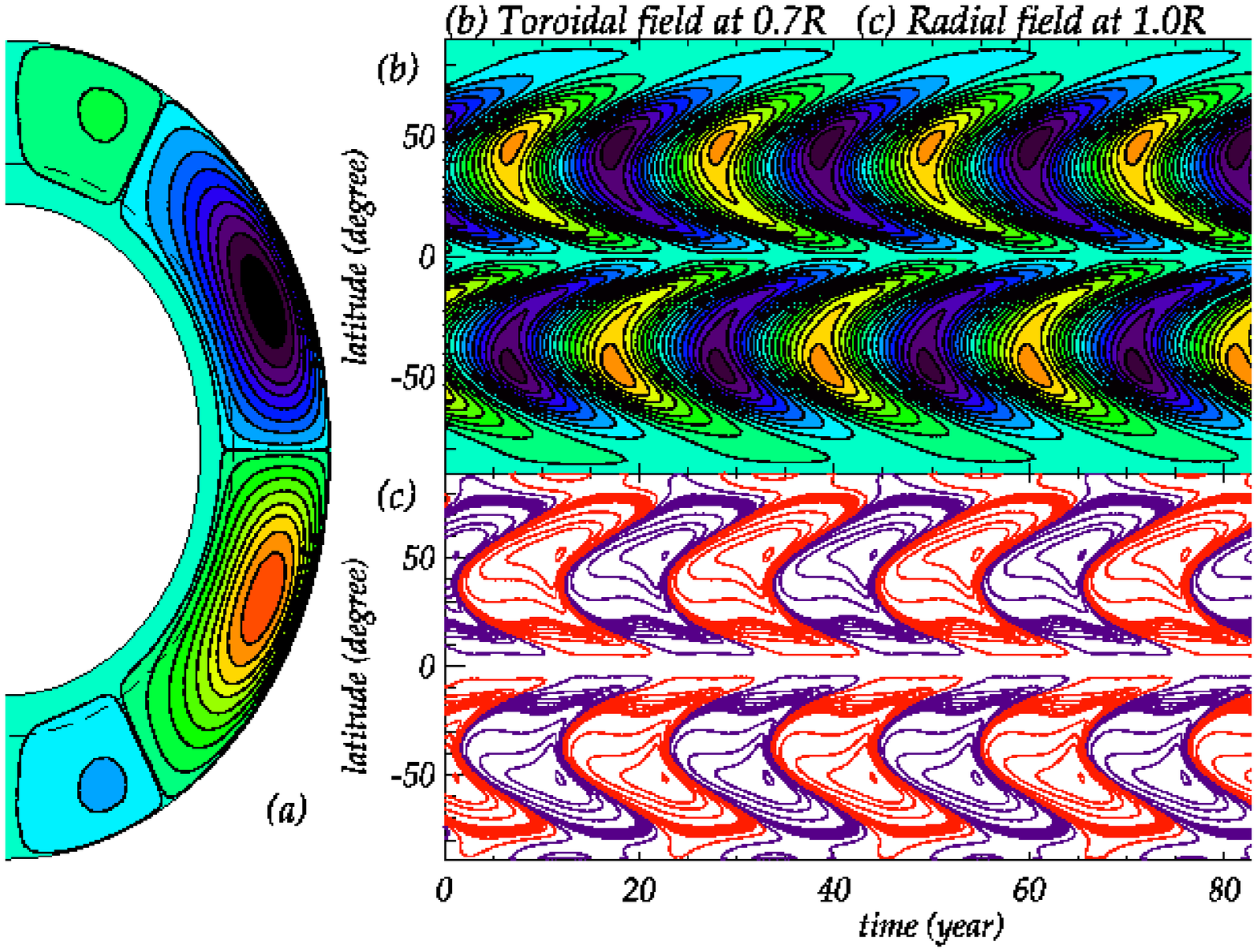}
\caption{Same as in Figure \ref{HHref} but the turbulent diffusivity is 
$\mathrm{7\cdot 10^{10}\,cm^2s^{-1}}$. The highest tachocline toroidal field 
is $\mathrm{\sim 78.\,kG}$ (red/violet). The maximum value of the radial 
fields is $\mathrm{\pm 77\,G}$, occurring near $50^{\circ}$.}
\label{HHloweta}
\end{figure}

Figure \ref{HHloweta} shows the time-latitude diagrams of toroidal and 
radial fields in the two cell case (Figure \ref{HHloweta}a) with 
$\mathrm{7\cdot 10^{10}\,cm^2s^{-1}}$ turbulent diffusivity. For this 
case we see that the high and low latitude branches of the butterfly 
diagram for tachocline toroidal field are about the same, despite the 
unequal amplitudes of the two circulation cells. Both the poleward 
meridional circulation at the bottom and the radial gradient of rotation 
in high latitudes at tachocline depths are contributing to this pattern, 
which is more pronounced than seen in Figure \ref{AAloweta}, for which the bottom 
meridional circulation is toward the equator at high latitudes. The sunspot
cycle is shorter, about $\mathrm{10.7}$ years, due to the shorter 
conveyor belt that is the primary circulation cell. Here too, the 
induced toroidal field are stronger than in the high diffusivity 
solution with the same meridional circulation. The maximum value of the 
toroidal field is $\mathrm{78\,kG}$; maximum radial field is 
$\mathrm{77\,G}$. In this case we get substantially less overlap 
of adjacent cycles than in the low diffusivity reference case (Figure 9).

\begin{figure}[hbt!]
\epsscale{1.0}
\plotone{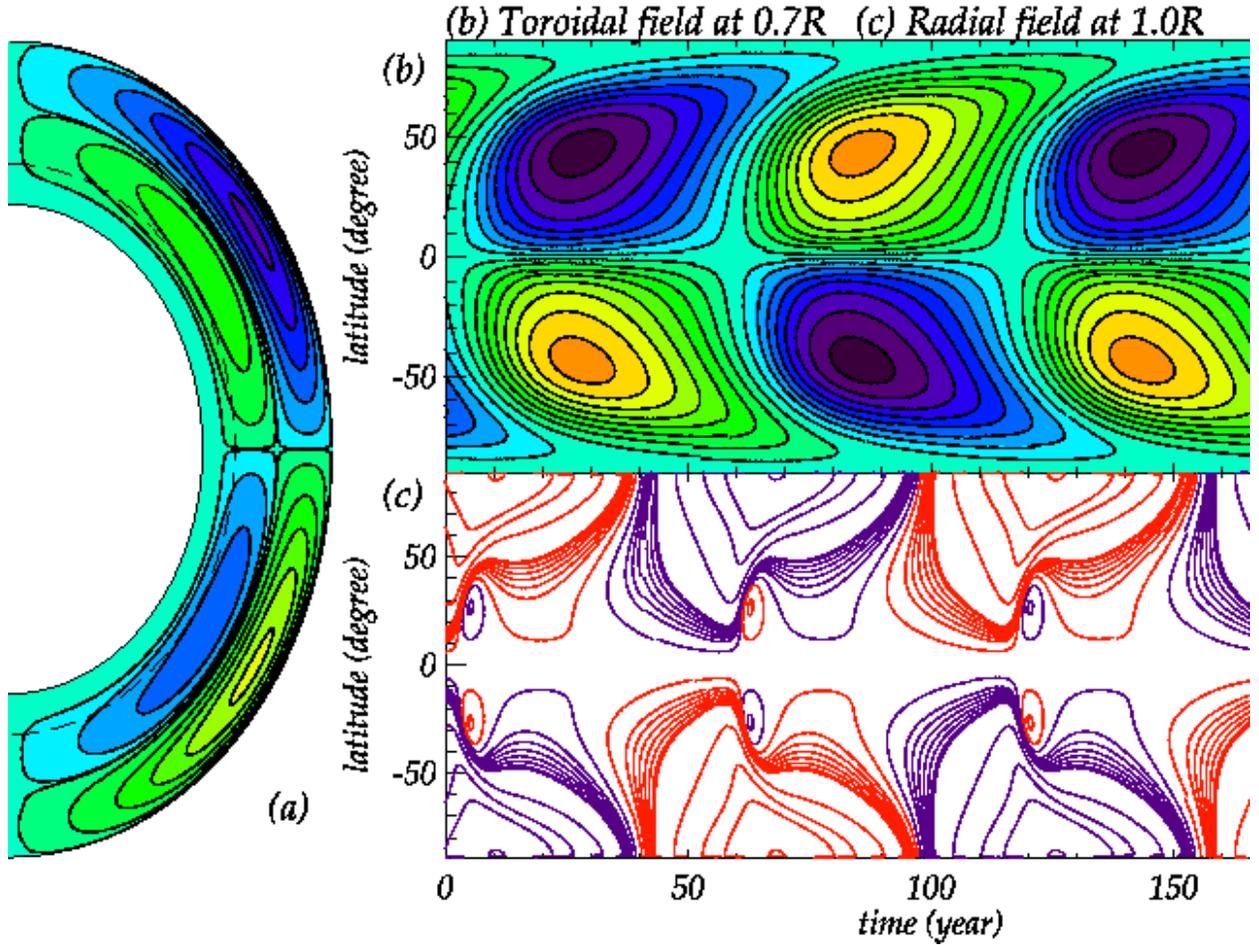}
\caption{
Same as in Figure \ref{SDOSDOref} but the turbulent diffusivity is 
$\mathrm{7\cdot 10^{10}\,cm^2s^{-1}}$. The highest tachocline toroidal 
field is $\mathrm{\sim 114\,kG}$ (red/violet). The maximal value of 
the radial fields is $\mathrm{\pm 288\,G}$, near the poles.
}
\label{SDOSDOloweta}
\end{figure}

\clearpage
In the next simulation with low diffusivity, we turn on a second meridional 
cell below the primary cell, as for the case previously displayed in Figure 
\ref{SDOSDOref}. The results are shown in Figure \ref{SDOSDOloweta}. 
The most prominent feature of the butterfly diagram is the longer 
sunspot cycle, about $\mathrm{50}$ years. What little migration of 
toroidal field there is, is toward the poles, leading to a slightly
antisolar butterfly. Here again, as in the higher diffusivity case shown 
in Figure \ref{SDOSDOcenterref}, toroidal field contours (not shown) 
at mid-depth of the convection zone, yields a butterfly diagram with 
both poleward and equatorward branches.

\clearpage

\begin{figure}[hbt!]
\epsscale{1.0}
\plotone{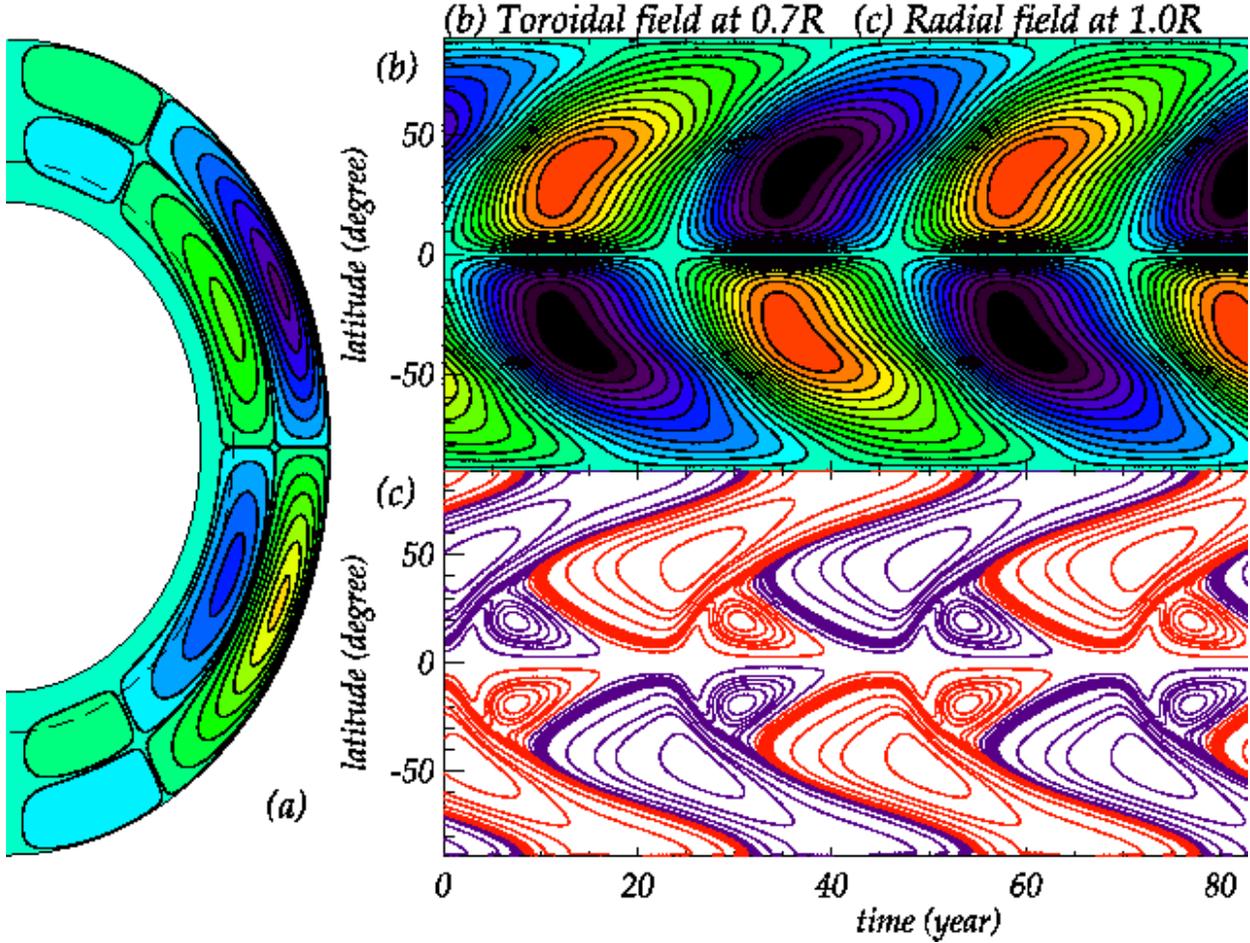}
\caption{ 
Same as in Figure \ref{SDOHHSDOref} but the turbulent diffusivity is 
$\mathrm{7\cdot 10^{10}\,cm^2s^{-1}}$. The highest tachocline toroidal 
field is $\mathrm{\sim 55\,kG}$ (red/violet). The maximum radial field 
is $\mathrm{\pm 83\,G}$, from about $50^{\circ}$ to near the poles.}
\label{SDOHHSDOloweta}
\end{figure}

The results of the next simulation is shown in Figure \ref{SDOHHSDOloweta}. 
Here, with two strong cells in low and mid-latitudes, and two reversed, 
relatively weak cells in polar latitudes, we again get a predominantly antisolar 
butterfly but with a small domain of equatorward migration of surface poloidal 
fields in low latitudes. This must be caused by the equatorward flow at mid-depth 
coupled with upward magnetic diffusion to the surface. Finally, here again we get 
a solar like butterfly for low and middle latitudes when we plot toroidal field 
contours at mid-depth in the dynamo domain.

\clearpage

\begin{figure}[hbt!]
\epsscale{1.0}
\plotone{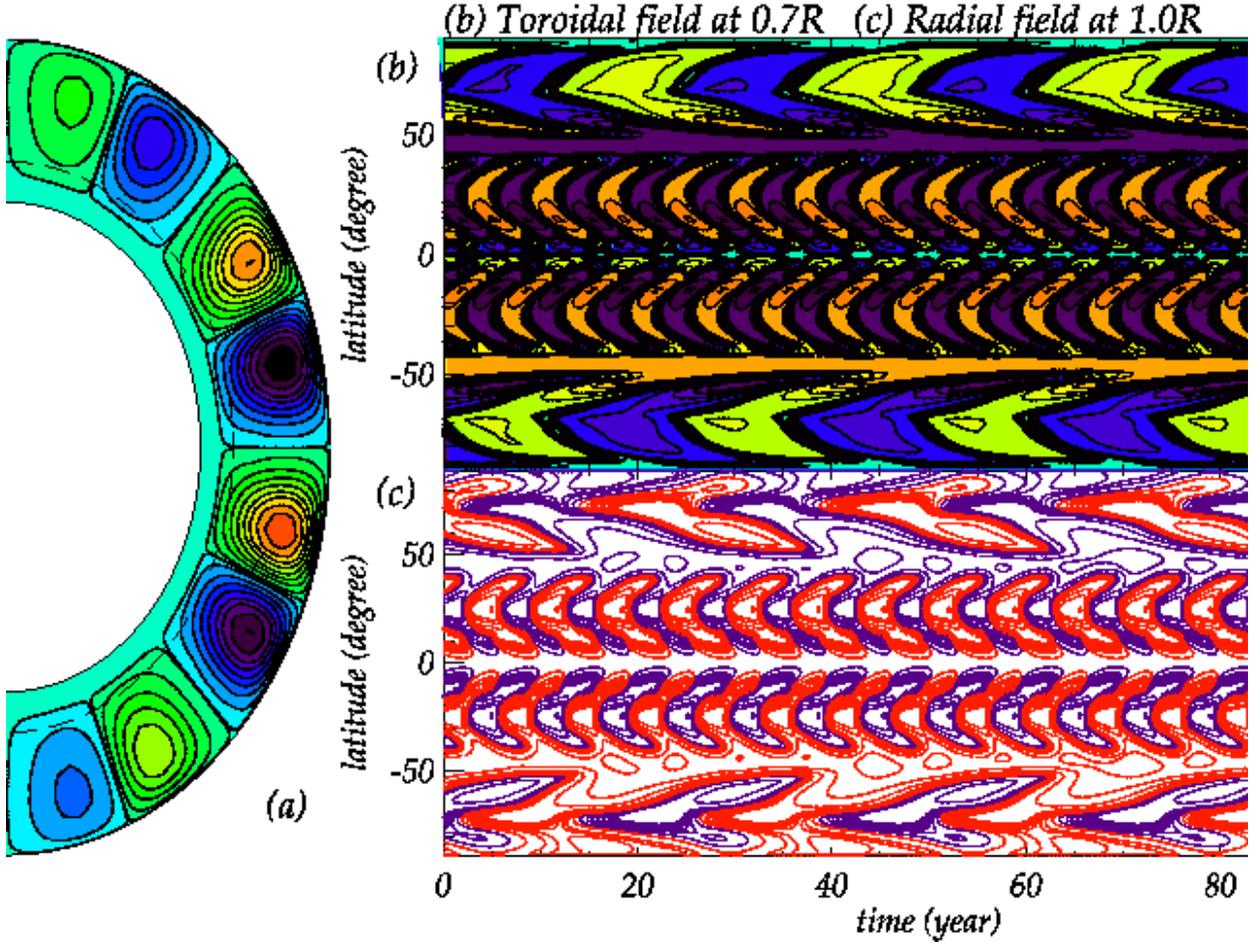}
\caption{Same as in Figure \ref{OOref} but the turbulent diffusivity is 
$\mathrm{7\cdot 10^{10}\,cm^2s^{-1}}$. The highest tachocline toroidal 
field is $\mathrm{\sim 65\,kG}$ (red/violet). The maximum value of the 
radial fields is $\mathrm{\pm 27\,G}$.}
\label{OOloweta}
\end{figure}

In the last simulation, we show results for the four cell case for 
low diffusivity in Figure \ref{OOloweta}. Due to the short conveyor belts, 
decreasing the turbulent diffusivity does not significantly change the 
cycle length. But the lower turbulent diffusivity has other effects. 
First, the fields are stronger, as we should expect. Second, we can 
see dynamo activity at the higher latitudes than in the high diffusivity 
case, though it is still low compared to that in low latitudes. The 
orientation of the wings of the time-latitude diagram at the different 
latitudes is determined by the latitudinal direction of the flow near 
the bottom; equatorward flow leads locally to migration toward the 
equator, and poleward flow to migration toward the poles, as seen in 
Figure \ref{OOloweta}b.

From all the time-latitude plots shown above, we can see that the most 
solar-like diagrams are produced if there is a single primary circulation
cell in each hemisphere, with possibly a weaker secondary, reversed cell
in polar latitudes. Circulation with two cells in depth, or two cells in
both latitude and depth, give solar-like butterflies only from toroidal
fields at mid-depth, not the bottom. For these to be correct for the Sun, 
the toroidal fields at the bottom must not come to the surface because
of their magnetic buoyancy or for any other reason, and a mechanism must 
exist that keeps mid-depth toroidal fields from rising buoyantly too fast 
to be amplified to produce sunspots. Neither requirement is easily satisfied
using known MHD processes.

\subsection{Parameter survey}

We have shown how solutions from a solar flux transport dynamo model 
differ for different forms of meridional circulation. The solutions 
we obtained are all for the same differential rotation, since from 
helioseismic measurements that is relatively well known for the Sun. 
These were found for fixed meridional circulation amplitude, but that 
amplitude is not very constrained from observations, and it is bound 
to have time variations. In addition, there are uncertainties in the 
amplitude and form of the surface poloidal flux source as well as the 
turbulent magnetic diffusivity. It is of interest to know how basic 
characteristics of a simulated sunspot cycle differ for different 
values of these uncertain parameters. Three prominent features of 
simulated cycles to focus on are its period, amplitude, and shape --
the times spent in ascending and descending phases. We focus here on
the first two of these.

We choose the range of meridional circulation amplitudes to cover
the range observed, and somewhat beyond. The choice of surface
poloidal source amplitude range is guided by estimates of relation
to photospheric magnetograms as discussed in \citet{dg06}. While
the choice of ranges of values for the meridional circulation amplitude
and surface poloidal source is guided by observations, and hence adequately
map the plausible parameter space for the Sun, the range of the
turbulent diffusivity amplitudes is selected based on the values 
that produce a sustaining dynamo. Given the turbulent diffusivity
is one of the biggest unknown ingredients in the solar interior, we
performed extensive numerical experiments to choose the range of $\eta_T$
so that the dynamo does not die due to too much diffusive decay,
or does not produce unusually large magnetic field amplitudes due to
too low diffusivity.  

\clearpage
\begin{figure}[hbt!]
\epsscale{0.52}
\plotone{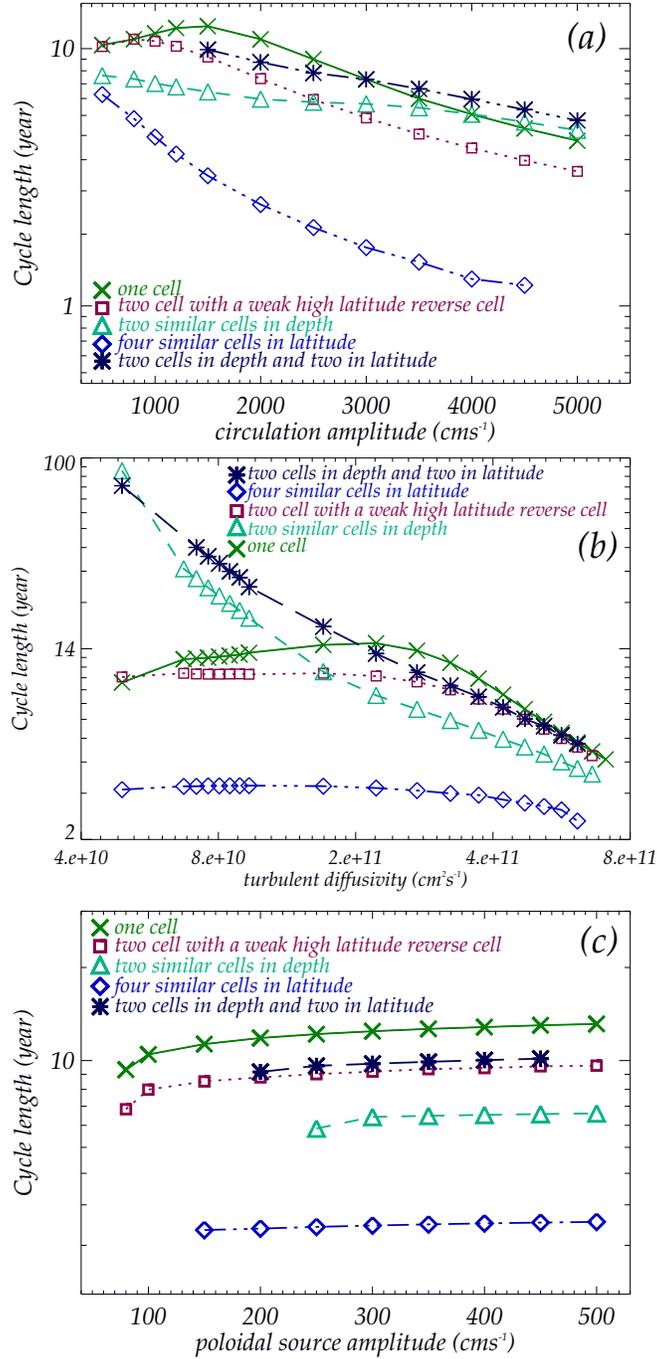}
\caption{Dependence of simulated sunspot cycle length on (a) amplitude 
of meridional circulation, (b) turbulent diffusivity and (c) amplitude 
of poloidal source term for the five circulation patterns used. The case 
of two cells in both latitude and depth is shown only for circulation 
amplitudes $\geq 15 {\rm m}{\rm s}^{-1}$. Below that amplitude, the 
solutions are quadrupolar rather than dipolar.}
\label{cyclelength}
\end{figure}
\clearpage

Figure \ref{cyclelength} displays the variation of cycle period with 
circulation amplitude (panel (a)), turbulent diffusivity (panel (b)) 
and poloidal source amplitude (panel (c)). We see from panel (a) that 
as the circulation amplitude is increased, in almost all cases the 
period declines. This is to be expected, because in all cases, unless 
diffusion dominates, the period is set by the speed of the conveyor belt. 
The primary exception we see is, that for low speeds, decreasing the 
circulation in the single cell case (and to much lesser degree, the case
with a second weak cell at high latitudes) leads to a decrease in period. 
This happens because, while the circulation is decreased, turbulent diffusion 
starts to 'short circuit' the conveyor belt, since some poloidal flux is 
diffused toward the bottom from the top before it reaches polar latitudes. 

This short circuiting effect is even more evident in panel (b), where we 
have plotted cycle length versus turbulent diffusivity. For the same 
circulation amplitude, the solutions become more diffusion affected to 
the right in the figure. The periods decline, in some cases by factors 
of five or more. This result shows that to have a flux transport dynamo 
calibrated to the observed sunspot cycle period requires careful choice 
of the turbulent diffusivity, no matter what circulation pattern is 
assumed. By contrast, panel (c) shows that the cycle period is almost 
independent of the amplitude of the surface poloidal source. This is also 
expected, because the dynamo is nearly linear. Changing the poloidal 
source amplitude should change primarily the peak amplitude of the
cycle, as we shall see below.

Figure \ref{maxfield} shows how the maximum toroidal field varies with meridional 
circulation (panel (a), turbulent diffusivity (panel (b)) and amplitude of the 
poloidal source term (panel (c)) for the five circulation patterns. From panel 
(c) we see that, as we should expect, raising the amplitude of the poloidal source 
raises the peak toroidal field amplitude. Because of the nonlinear quenching of 
the source term internal to the model, however, the amplitudes are beginning to 
approach asymptotic limits. There are also significant differences in the
efficiency of different circulation patterns; two cells in depth and four cells 
in latitude both generate much less toroidal field than the other patterns, 
which give almost the same amplitudes. In the four cell case, this is because 
the shearing of poloidal field is largely confined to low latitudes, where the 
latitudinal rotation gradient is weakest, by the short latitudinal extent of 
the conveyor belt, so less toroidal field is generated. In the case of two 
cells in depth, less of the poloidal flux gets to the bottom where the radial 
shear is strongest, again reducing the dynamo's ability to amplify toroidal 
field.

\clearpage
\begin{figure}[hbt!]
\epsscale{0.52}
\plotone{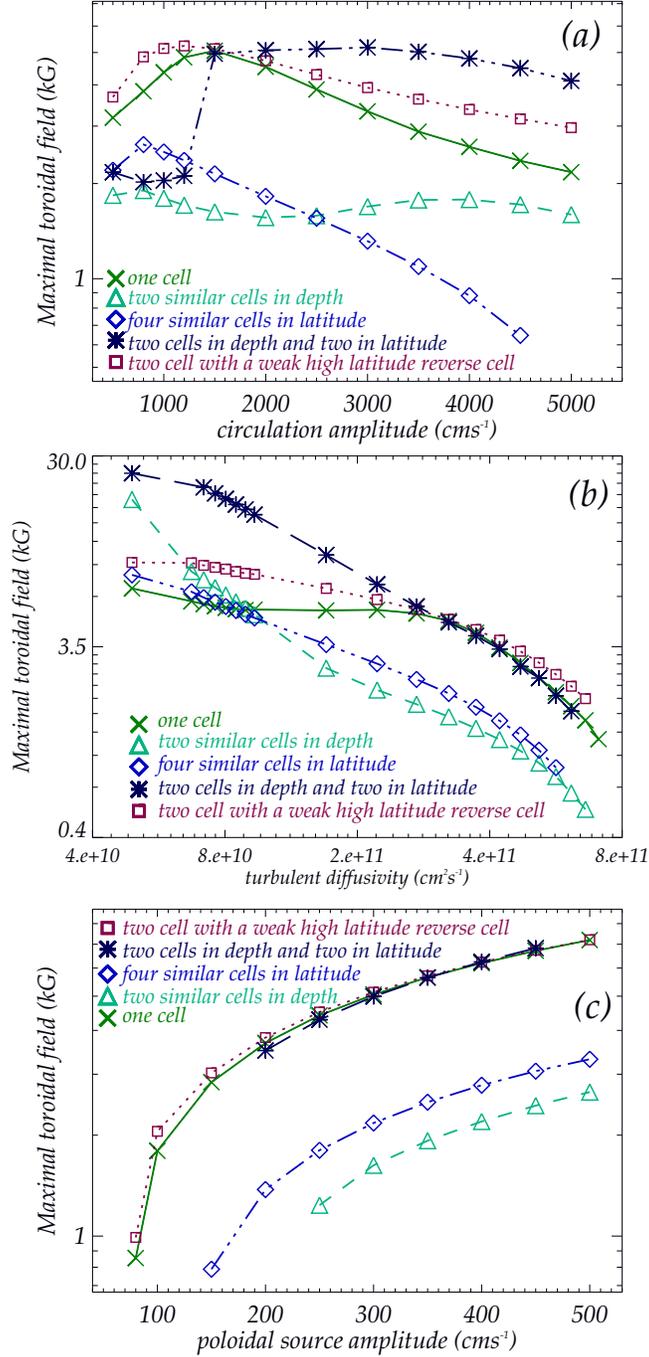}
\caption{Dependence of maximum toroidal field on (a) meridional circulation 
amplitude, (b) turbulent diffusivity and (c) amplitude of poloidal source 
term, for the five circulation patterns used. The sudden drop in cycle
amplitude for the case of two cells in latitude and depth is due to the 
solution switching from dipolar to quadrupolar at a peak circulation 
amplitude of $15 {\rm m}{\rm s}^{-1}$.}

\label{maxfield}
\end{figure} 
\clearpage

In panel (b) the result that peak toroidal field declines with increasing 
magnetic diffusivity is due to the fact that there is more dissipation in the 
system that the induction processes in the dynamo have to overcome. Here too 
we see that, for all diffusivities, configurations with four cells in latitude 
and two cells in depth generate less toroidal field for the same assumed 
diffusivity.

From panel (a) we see that the variation in peak toroidal field with changes in 
circulation amplitude are more complex. With two cells in depth the peak amplitude 
changes relatively little with circulation amplitude. With two cells in both 
latitude and depth, we see a similar result, except near circulation amplitude 
$15 {\rm m}{\rm s}^{-1}$ where the amplitude drops by more than $50\%$ when the 
magnetic field configuration switches from dipolar to quadrupolar. In both cases, 
with the relatively fine scale meridional circulation pattern, the solutions are 
in the diffusivity dominated regime, so the amplitude does not change. Meridional 
circulation acts mainly as a transporter of toroidal and poloidal flux, rather 
than as an amplifier. The faster the toroidal field is transported, the weaker
is the resulting field amplitude, because the toroidal field peak is displaced 
before it amplified as much as it could be if it were stationary.

The other three cases each show a largest value of toroidal field amplitude at 
a meridional circulation amplitude between about $8 {\rm m}{\rm s}^{-1}$ and 
$15 {\rm m}{\rm s}^{-1}$. At these speeds there is an optimum balance between 
amplification of toroidal field by differential rotation shearing, diffusive 
decay, and meridional transport of toroidal and poloidal flux. In each case, 
for larger than optimum circulation, the toroidal fields are moved in latitude 
and/or depth too fast to be as fully amplified as it would have been if moving
more slowly; at less than optimum advection rate, more time is allowed at a 
given latitude and depth for the toroidal and poloidal fields to decay due
to diffusion.

\citet{jb07} computed a power law relationship among the various parameters
varied for the case of two cells in latitude and two in radius. However, for
variation in flow structures in terms of not only number of cells in latitude
and depth, but also the node locations in latitude and depth, it will not be
possible to reach a unique answer for the power law relationship. Thus we
presented here the curves as function of flow-speed, diffusivity and poloidal
source amplitudes. Furthermore, some of these flow patterns do not produce
solar-like features, and hence power law relationships may not be meaningful.

\subsection{Parity issue}

In all the cases we have presented so far, we have found dipolar parity
during the simulation time of interest, i.e. about up to 500 years.
We know that the growth rates of quadrupolar parity solutions are
slightly higher than the dipolar parity in a Babcock-Leighton dynamo 
and the dipolar parity slowly drifts to quadrupolar one if the dynamo 
simulations run for more than 2000 years \citep{dg01, berb02, hy10, md14}. 
However, \citet{jb07} have shown that this switching from dipolar to 
quadrupolar parity is very fast in the case of a four-celled meridional 
circulation that consists of two cells in latitude and two in depth. In our 
four-celled case consisting of two cells in latitude and two in depth (see 
Figure \ref{SDOHHSDOref}), the parity change did not occur so quickly. This 
is because the ratio of poleward surface flow-speed of the top cell to that 
of the bottom cell was too high, $\sim 50$, whereas in \citet{jb07}
that ratio was $\sim 6$. In order to investigate the fast change of
parity in the four-celled case, we consider a ratio of poleward
surface flow-speed to poleward bottom flow-speed to be $\sim 5$ and
simulate that case and present our results in Figures 16-19. 

Even though the four-celled case with two cells in latitude and two 
in depth produce least solar-like solutions, it is worth examining the
parity issue, because such a four-celled profile could appear intermittently
in one or both hemispheres of the Sun. Fast change in parity in one hemisphere
with respect to the other could lead to a large phase shift between North 
and South cycles, as observed in recent cycle.

In Figure 16, as in earlier figures panel (b) shows the toroidal fields 
near the bottom of convective zone. Panel (c) depicts the surface radial field. 
We see here a radical change from the earlier examples. The toroidal 
and poloidal field patterns are now symmetric rather than antisymmetric 
about the equator. In other words, we have found quadrupolar type rather 
than dipole type parity. This difference in parity about the 
equator develops in just a few cycles, so the system in this case has a 
strong preference for quadrupole parity. 

\citet{jb07} demonstrated, in their simulations of dynamo with four-celled
(two in latitude and two in depth) meridional circulation, for what 
combinations of meridional flow speed and turbulent magnetic diffusivity 
quadrupolar structure is favored. In the case of a meridional circulation 
consisting of single cell in each hemisphere, \citet{dg01} reasoned that 
slow switching to quadrupolar from dipolar parity occurs when bottom 
poloidal fields become weak enough after a long traversal via the 
conveyor-belt from surface to the bottom, and hence cannot connect with 
their opposite-hemisphere counterparts about the equator. A global 
statement would be that the quadrupolar mode is selected when, for a 
particular meridional circulation, it is dissipated at a substantially 
lower rate than is the dipolar mode. In other words, the growth rate 
for a quadrupolar mode is higher than for a dipolar mode. Since a 
circulation pattern that has two cells each in both latitude and 
depth is inherently more complicated than a simple, single-celled pattern,  
the poloidal and toroidal flux of opposite signs and different amplitude 
are converged together in more places away from the equator. So the
dynamo favors quadrupolar symmetry, unless the upper and lower cells 
are very unequal in amplitude.


\begin{figure}[hbt!]
\epsscale{1.0}
\plotone{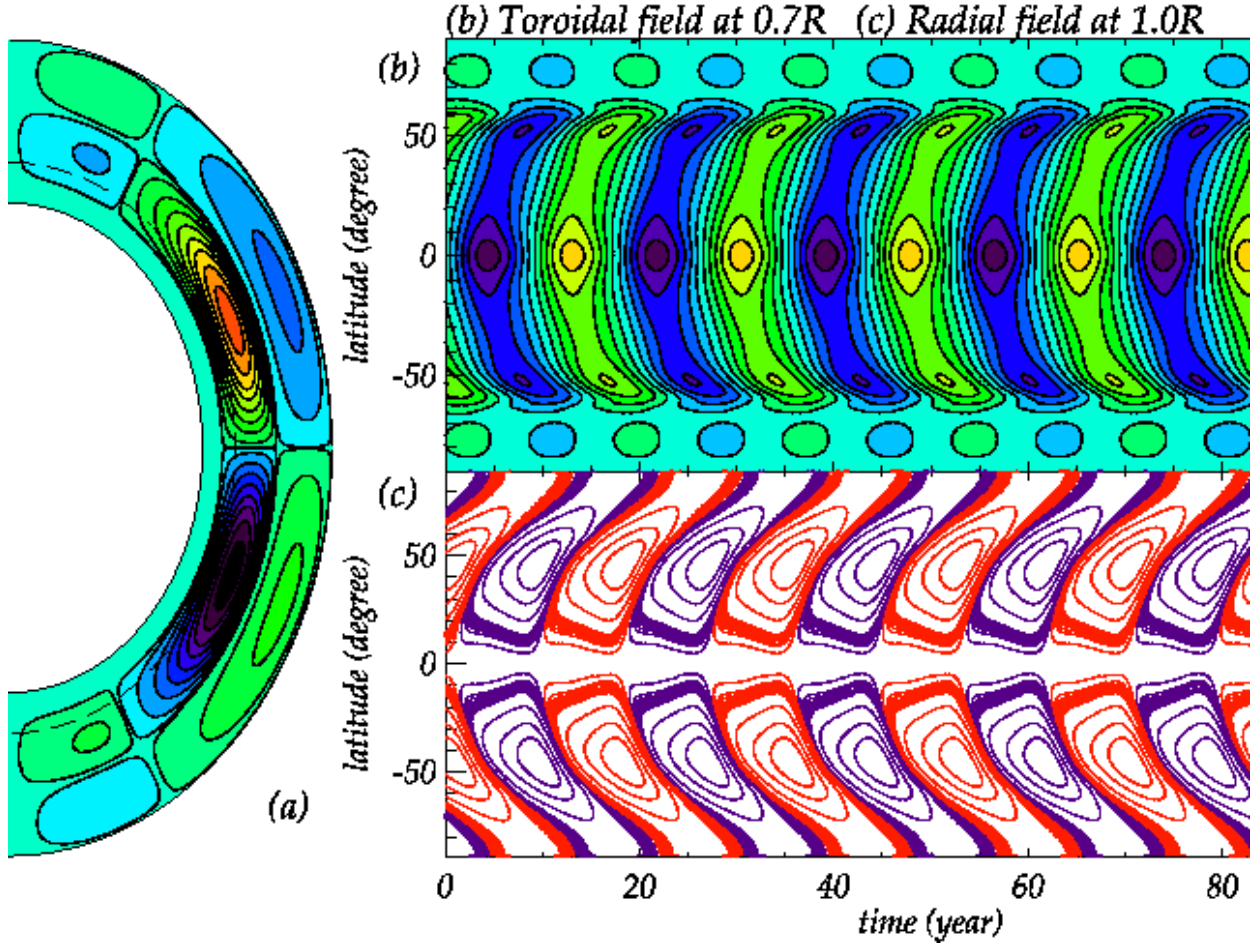}
\caption{Same as in Figure \ref{SDOHHSDOref} but the bottom cell of
this four-celled meridional circulation pattern has about ten times 
stronger poleward flow compared to that in \ref{SDOHHSDOref}a.
The highest toroidal field amplitude is about $\mathrm{\sim 27\,kG}$ 
(yellow/violet). The maximum value of the radial fields is 
$\mathrm{\pm 150\,G}$, near $50^{\circ}$}.
\label{SDOHHSDOref_alt}
\end{figure}

We find also in the low diffusivity case that having two cells in 
both latitude and radius leads to a quadrupolar solution, if the ratio
of surface-flow to bottom-flow is 5:1 instead of 50:1 (we do
not produce here the time-latitude diagram for the low diffusivity
case). Therefore this fast switching from dipolar to quadrupolar parity 
is a typical phenomena in the four-celled meridional circulation with
a much stronger bottom-flow of similar order of magnitude as the surface-flow, 
irrespective of turbulent diffusivity value.

Figures 17 and 18 give details of the actual transition from dipolar to 
quadrupolar symmetry. In Figure 17, panels (a) and (c) we see by eye
that the transition appears to occur in about 120 years (16-18 sunspot 
cycles) in the tachocline toroidal field and surface poloidal field 
respectively. Frames (b) and (d) show finer detail for each in the 
middle of this transition. The simulation was started from a previous 
dipole simulation, which inevitably has some slight departures from
dipole symmetry at the truncation error level. The quadrupolar symmetry is 
so strongly preferred for the parameters chosen that even these small
differences are enough to start the process of symmetry switching.

What we see in Figure 17 is a very simple process in which switching
occurs by the northern hemisphere developing a phase lag relative to the 
southern hemisphere, which grows until the North lags by a sunspot or half
magnetic cycle, with very little change in pattern in each hemisphere.
What is actually happening is that the periods of both hemispheres
are getting longer, from about 6.5yr to 8.5yr, but the North reaches the longer 
period faster than the South, so the South gains on it in phase. 
Presumably in another simulation with different truncation errors, it could
be the northern hemisphere that lags, ending up with the same final state.

In Figure 18, we show meridional cross-sections of both toroidal field
(blue/yellow shading) and poloidal field (solid black and dashed red lines)
in 1-year intervals in the middle of the transition. We can see particularly 
that the peak toroidal fields are moving up in the domain to a mid-depth 
where the flow in low latitudes is toward the equator. In addition, the polar 
field reversals are evolving to a state in which they go from positive to 
negative at both poles at the same time.

\clearpage
\begin{figure}[hbt!]
\epsscale{0.9}
\plotone{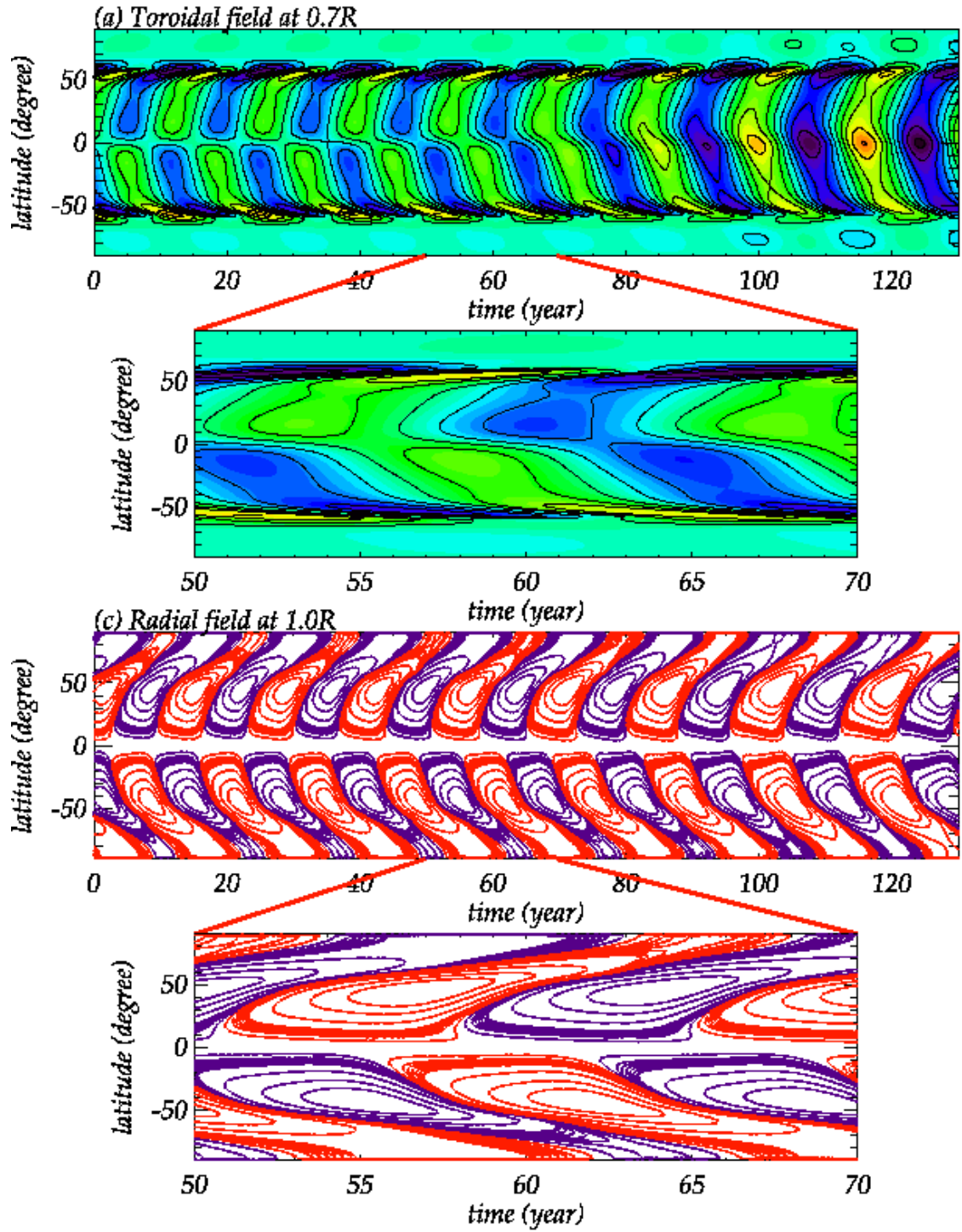}
\caption{Panels (a) and (c) show time-latitude diagrams for tachocline 
toroidal fields and surface radial fields respectively; evolution of 
parity from dipolar to quadrupolar is shown in enlarged form in panels
(b) and (d).} 
\label{paritychange}
\end{figure} 

\begin{figure}[hbt!]
\epsscale{0.85}
\plotone{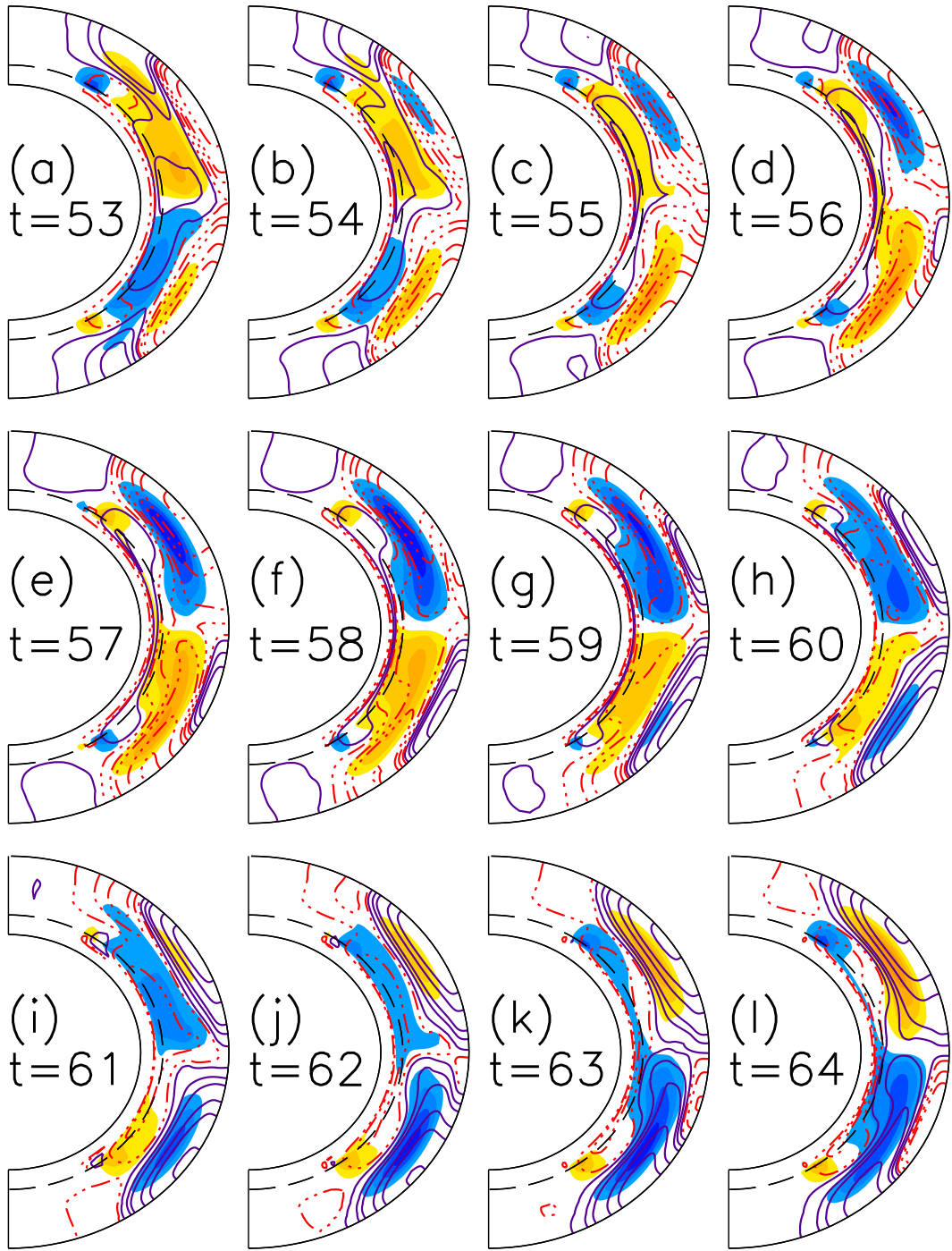}
\caption{Twelve snapshots of toroidal fields in orange/violet color-filled
contours and poloidal fields in red (positive) and blue (negative) 
contours show the evolution of fields during parity change.}
\label{field_parity_evolution}
\end{figure} 
\clearpage

We can quantify how long the transition takes by measuring the difference
in cycle periods between South ($P_S$) and North ($P_N$) as a function 
of time. The normalized difference in cycle periods between South and
North $(P_S - P_N)/[(P_S+P_N)/2]$ is shown as function of time in 
Figure 19. We see that during this transition, the North adjusts to 
the longer period faster than the South. The difference in cycle period 
is positive, and reaches a peak of about $12\%$ midway through the 
transition.

\begin{figure}[hbt!]
\epsscale{1.0}
\plotone{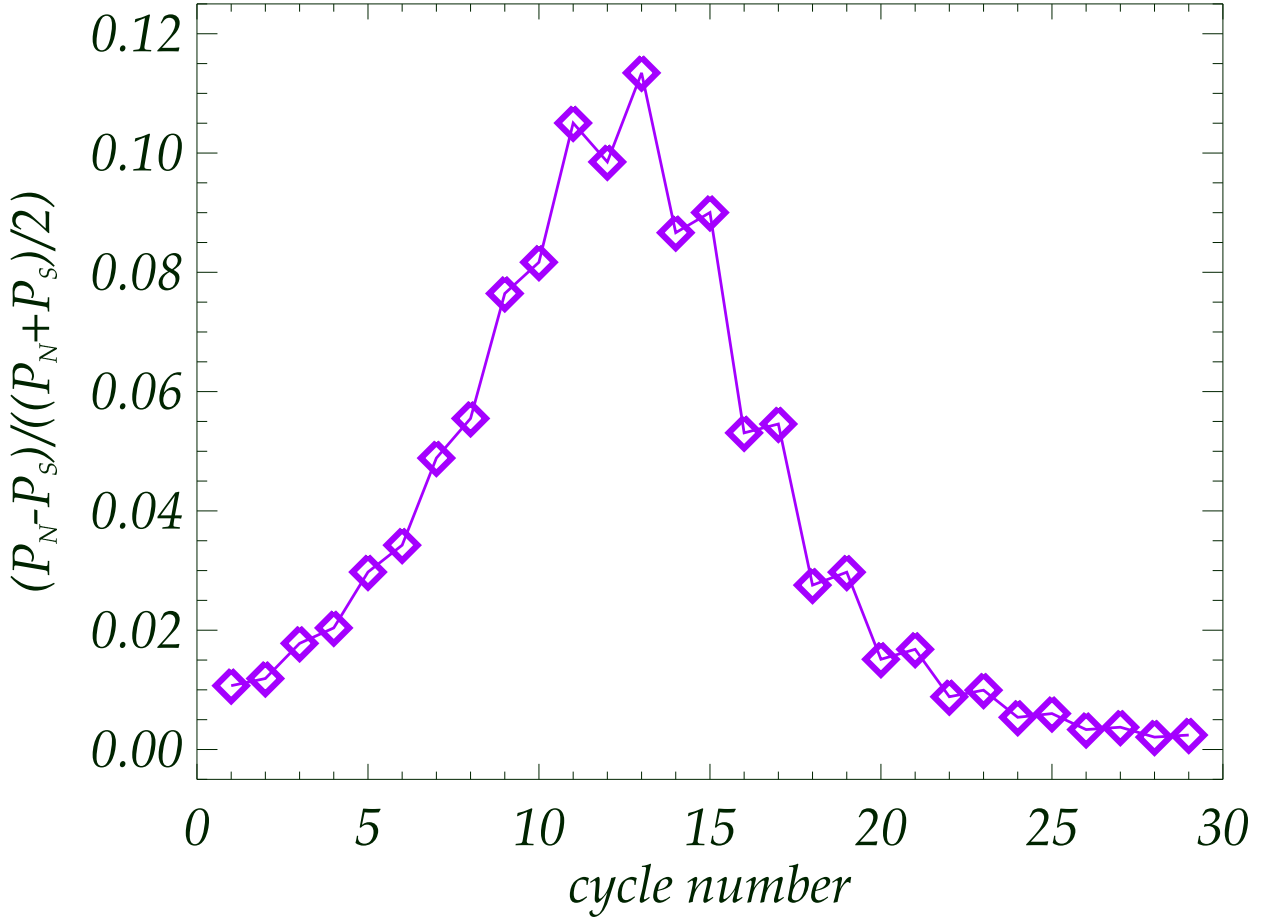}
\caption{Normalized difference in cycle-lengths in North and South as
function of cycle number during change in dipolar to quadrupolar
parity.}
\label{cyclelength_evolution}
\end{figure} 

This switching is possible because with the circulation pattern chosen, there 
is much less linking of flux between hemispheres at the equator. With dipole 
symmetry there is strong diffusion across the equator, whereas with quadrupole 
symmetry there is much less since both sides have the same sign of field there. 
Furthermore, dipole symmetry is best preserved when the toroidal fields
of both signs are strongest where they are being brought into close 
proximity near the equator. This is guaranteed when there is single cell 
in depth in low and mid latitudes, with equatorward flow near the bottom where 
the turbulent magnetic diffusion is smaller.

The parity switching example we have shown above is very different from that
found in \citet{jb07}, particularly in latitudes equatorward of $45^{\circ}$.
This is not surprising, because there is a significant structure in the flow
structures -- we have two strong primary cells at the top and bottom, extending 
from the equator to $60^{\circ}$ latitudes, associated with two weak secondary 
cells at polar regions, whereas \citet{jb07} had four equally strong cells 
(two in top and two in bottom) of equal latitudinal extents. We found a simple 
drift of phase in one hemisphere with respect to the other at all latitudes, 
while in \citet{jb07} (Figure 10) there is a radical rearranging of poloidal 
flux in low latitudes, with a switch in symmetries by the antisymmetric mode 
dying out, replaced by the symmetric mode without much change in phase locally. 
This difference suggests there may be multiple ways for symmetry switching to 
occur, which should be explored in the future.

\section{SUMMARY AND CONCLUSIONS}

We have compared flux-transport dynamo model results for five meridional
circulation patterns that may occur in the solar convection zone, as
suggested by solar observations and/or hydrodynamic models and full
3D simulations applied to the Sun. We carried out simulations for both 
diffusion and advection dominated regimes. Only the circulation pattern 
is different in each simulation; all other physical processes included 
are the same. We find a wide variety of dynamo behavior, as measured by 
simulated time-latitude diagrams of toroidal and poloidal fields. 

In general, circulation patterns with only one cell in depth and no more 
than two cells in latitude produce the most solar like butterfly diagrams. 
Two cells in depth leads to antisolar butterflies from tachocline toroidal 
fields, but solar like butterflies at mid-depth where both cells have 
equatorward flow. For this pattern to work for the Sun physical mechanisms 
must exist to inhibit magnetic buoyancy there long enough to allow enough 
amplification of toroidal fields to produce spots, while preventing 
tachocline toroidal fields from reaching the solar surface in any 
observable form. Four cells in latitude leads to some solar-like magnetic 
patterns, but very fast cycle periods compared to the Sun. Surface Doppler 
measurements also do not support the existence of four cells distributed 
evenly in latitude, though multiple cells confined to polar latitudes 
can not be ruled out.

All of the solutions we have found retain dipole or solar-like symmetry
about the equator within the simulation span of 500 years, except the 
case of circulation pattern with two cells in latitude and and two in
depth when the flow speeds in the upper and lower cells differ by less 
than a certain amount. In that case, from a small difference between 
hemispheres starting probably from the numerical truncation, the solution 
switches to quadrupole type within several magnetic cycles and stays 
there for ever, even when starting from essentially dipole symmetry. 
This switch is achieved simply by one hemisphere temporarily changing 
its period relative to the other until the relative phase changes by 
one-half cycle, without changing the pattern itself in either hemisphere. 
A milder version of this effect could be partly responsible in the Sun 
for differences in phase between northern and southern hemispheres that 
do not go so far as to switch the dominant symmetry observed, which is 
dipolar.

Despite producing significantly different butterfly diagrams for toroidal
and poloidal fields, our flux transport dynamo simulations with different
meridional circulations have many properties in common, as revealed by our 
parameter survey. In almost all cases for all parameters chosen, cycle length
monotonically declines with increasing circulation amplitude and increasing 
turbulent magnetic diffusivity, but is nearly independent of poloidal 
source amplitude. Maximum fields generated also decline monotonically with 
increasing diffusivity, but increase with poloidal source amplitude. However, 
changes in circulation amplitude produce non-monotonic changes in peak 
fields for different circulation patterns.

Throughout the calculation we have fixed the differential rotation profile
given by expressions (11) and (12). A differential rotation profile
that more accurately captures high-latitude pattern beyond $60^{\circ}$ and 
the near-surface shear layer requires additional terms. Performing simulations 
with such a differential rotation profile as given by \citet{schou98} (see 
also \citet{dctg02}), we did not find changes in dynamo cycle period or in
surface radial fields, but a small increase ($\sim 1.07\%$) in the tachocline 
toroidal field amplitude with respect to that obtained in the present paper.

There are at least two important effects related to MHD turbulence that 
we have not included in the model we have used that we need to examine in 
future studies. Both would add nonlinearities to the system that are
currently beyond the scope of formulation in this paper. One is the 
so-called 'turbulent pumping' mechanism \citep{tbct98, kkt06, gdg08} 
and references therein), and the other is diffusivity 'quenching' 
\citep{gddg09}. Apparently the turbulent pumping might be able 
to counter the magnetic buoyancy effect to keep the fields in the 
convection zone, especially for the solutions where the mid convection 
zone toroidal field shows more solar-like butterfly pattern for a meridional
circulation with two cells in depth. This effect should be explored in 
the future.

\acknowledgements

We thank Yuhong Fan for a thorough review of our manuscript. We extend
our thanks to an anonymous reviewer for his/her helpful and constructive 
comments on an earlier version of our manuscript, which have helped significantly
improve the paper. This work is partially supported by the Hungarian 
Science Research Fund with OTKA grant through the award number K83133, the 
Graduate Study Program of ASP at University Corporation for Atmospheric 
Research, under the contract number P-1-01560, and by NASA's Living With a 
Star grant through the award number NNX08AQ34G. The National Center for 
Atmospheric Research is sponsored by the National Science Foundation.

\end{document}